\documentclass[compsoc]{IEEEtran}
\usepackage{amsmath,mathtools}
\usepackage{amsthm}
\usepackage{graphicx}
\usepackage{multirow}
\usepackage{color}
\theoremstyle{definition}

\usepackage[linesnumbered,algoruled,noline,noend]{algorithm2e}
\theoremstyle{definition}

\usepackage{enumitem}
\usepackage{array}
\newcolumntype{L}[1]{>{\raggedright\let\newline\\\arraybackslash\hspace{0pt}}m{#1}}
\newcolumntype{C}[1]{>{\centering\let\newline\\\arraybackslash\hspace{0pt}}m{#1}}
\newcolumntype{R}[1]{>{\raggedleft\let\newline\\\arraybackslash\hspace{0pt}}m{#1}}

\definecolor{grey}{rgb}{0.6,0.6,0.6}

\newcommand{\rev}[1]{{\textcolor{blue}{#1}}}

\usepackage{adjustbox}
\usepackage{listings}
\usepackage{xspace}
\usepackage{xfp}
\usepackage{balance}
\usepackage{soul}
\usepackage{tikz}
\usepackage{calc}
\usepackage{makecell}
\usepackage{diagbox}
\usepackage{pifont}
\usepackage{framed}
\usepackage{marginnote}
\usepackage{threeparttable}

\colorlet{shadecolor}{gray!20}

\begin{document}

\twocolumn




\title{HTAP Databases: A Survey}

\author{Chao Zhang, Guoliang Li*, \textit{Fellow}, \textit{IEEE}, Jintao Zhang, Xinning Zhang, Jianhua Feng}

\IEEEtitleabstractindextext{%
\begin{abstract}
Since Gartner coined the term, Hybrid Transactional and Analytical Processing (HTAP), numerous HTAP databases have been proposed to combine transactions with analytics in order to enable real-time data analytics for various data-intensive applications. HTAP databases typically process the mixed workloads of transactions and analytical queries in a unified system by leveraging both a row store and a column store. As there are different storage architectures and processing techniques to satisfy various requirements of diverse applications, it is critical to summarize the pros and cons of these key techniques. This paper offers a comprehensive survey of HTAP databases. We mainly classify state-of-the-art HTAP databases according to four storage architectures: (a) Primary Row Store and In-Memory Column Store; (b) Distributed Row Store and Column Store Replica; (c) Primary Row Store and Distributed In-Memory Column Store; and (d) Primary Column Store and Delta Row Store. We then review the key techniques in HTAP databases, including hybrid workload processing, data organization, data synchronization, query optimization, and resource scheduling. We also discuss existing HTAP benchmarks. Finally, we provide the research challenges and opportunities for HTAP techniques.
\end{abstract}
\pagenumbering{arabic}

}
\setcounter{figure}{0}
\maketitle

\IEEEdisplaynontitleabstractindextext
\IEEEpeerreviewmaketitle

\section{Introduction}

\IEEEPARstart{H}ybrid Transactional, and Analytical Processing (HTAP) was defined by Gartner~\cite{Gartner2014htap, gartner2018}. Since then, HTAP techniques have been deployed in various data-intensive applications, e.g., banking and finance~\cite{Banking_HTAP}, E-commerce~\cite{OceanBase}, and fraud detection~\cite{FraudDetection1}. The Gartner report envisioned that, by 2024, HTAP techniques will be widely adopted in numerous business applications that entail real-time data analytics. Compared with the traditional data processing pipeline that processes transactions and analytical queries separately, HTAP architecture enables a unified system that not only can handle online transactional processing (OLTP) efficiently, but also can perform online analytical processing (OLAP) concurrently~\cite{gartner2018}. Such an architecture aims to eliminate the need for an explicit Extract-Transform-Load (ETL) process, thereby enabling real-time data analytics on transaction data. For instance, HTAP databases allow entrepreneurs in retail applications to analyze the latest transaction data in real-time, and then they can roll out promotional sales based on the gained insights. In finance applications, HTAP databases enable vendors to process customer transactions efficiently while detecting fraudulent transactions simultaneously~\cite{Application2016}.

Over the last decade, we have witnessed the emergence and evolvement of various HTAP databases \cite{Polynesia_ICDE2022, ByteHTAP, cubukcu2021citus, saphana2015, TiDB2020, Oracle2015, SQLServer2015, Greenplum2021,MariaDB2021,MySQLHeatWave2021,Hyper_column, Polardb2021, SingleStore2021,DB2BLU2013, SAPHANA2012, OceanBase}. Since it is well recognized that a row store is ideal for OLTP workloads, and a column store is better suited for OLAP workloads \cite{hybridindex2018, Oracle2015}, HTAP databases mainly adopt a dual-store architecture that leverages both a row store and a column store. However, different categories of HTAP databases adopt disparate storage strategies and processing techniques to cater to various applications. For instance, it depends on whether OLTP or OLAP has a higher priority for the applications (e.g., OLTP is the first citizen for banking scenarios while OLAP dominates the analytical reporting). It also depends on the requirements of availability, scalability, performance, and data freshness~\cite{datafreshness2004} specified in the service level agreements (SLAs)~\cite{saphana2015} (e.g., large-scale E-commerce applications require high scalability while banking and finance applications demand high data freshness). Nevertheless, HTAP databases must strike a trade-off between performance isolation and data freshness when handling the mixed workloads of OLTP and OLAP. The main reason is that the two types of workloads exhibit a completely different computing pattern and can intervene with each other, i.e., OLTP workloads \cite{sigmod/OLTP} are update-heavy and short-lived while OLAP workloads \cite{pvldb/OLAP} are read-heavy and bandwidth-intensive. Consequently, HTAP databases cannot guarantee the metrics of query throughput and data freshness simultaneously, albeit with a unified architecture. As different applications deliver disparate requirements, it is critical to study and understand the pros and cons of HTAP databases under various architectures.

\begin{figure}[!t]
	\centering
	\includegraphics[width=0.95\linewidth]{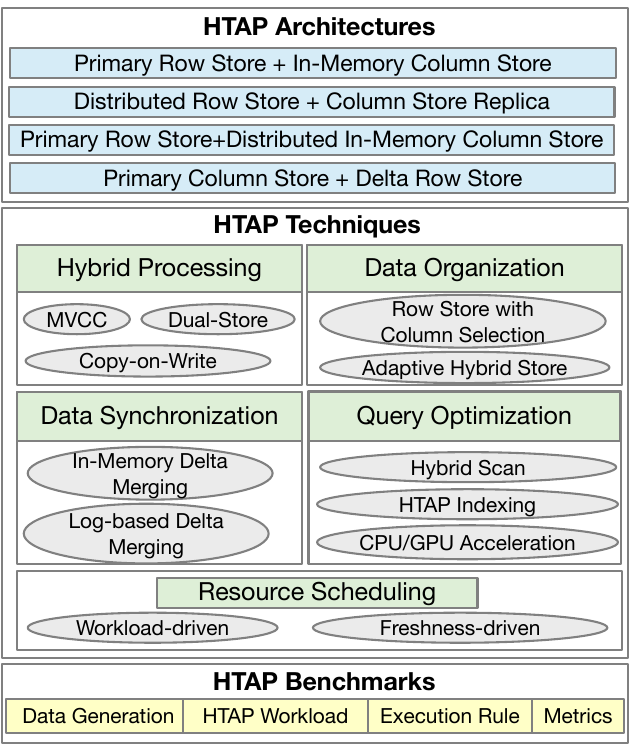} 
	\caption{An Overview of HTAP Architectures, Techniques, Benchmarks}	\label{fig:overview}
\end{figure}

There are five main challenges that HTAP databases need to address, including (C1) hybrid workload processing, (C2) data organization, (C3) data synchronization, (C4) query optimization, and (C5) resource scheduling.

\textbf{Challenge 1. Hybrid  Processing Challenge}. The first challenge is to process mixed workloads of OLTP and OLAP efficiently while maintaining a high data freshness. As there is a trade-off between performance isolation and data freshness, it is challenging for HTAP databases to balance performance and freshness. Hence, these systems must make a compromise on either performance isolation or data freshness, depending on the specific HTAP applications.

\textbf{Challenge 2. Data Organization Challenge}. The second challenge is to organize the data adaptively for HTAP workloads with high throughput and low storage costs. There are two rules of thumb: (i) the row store is suitable for OLTP workloads and (ii) the column store is ideal for OLAP queries. However, storing a copy of the entire data for both formats leads to high storage overhead (i.e., space cost and maintenance cost). Therefore, HTAP systems need to make wise decisions for data organization (e.g., row-wise or column-wise data layout) to reduce the storage overhead while delivering high system performance.

\textbf{Challenge 3. Data Synchronization Challenge.} The third challenge is to decide when to synchronize the delta data from the transactional store to the analytical store for high data freshness while keeping high throughput and scalability. On the one hand, immediately merging the delta data to the analytical store can keep high data freshness. However, it can greatly affect the performance due to the merging overhead. On the other hand, merging the delta data on demand can improve the throughput but lead to low data freshness. Hence, HTAP databases need to synchronize the data adaptively.


\textbf{Challenge 4. Query Optimization Challenge}. The fourth challenge is to optimize the queries with both a row store and a column store. In the HTAP databases, a query can be executed against either the row store or the column store; thus the query optimizer must judiciously decide whether the row-based execution or the columnar execution is more beneficial. However, it is challenging to generate an optimal query plan in 
a large search space. Therefore, the HTAP optimizer must balance the trade-off between the planning time and query execution latency. 


\textbf{Challenge 5. Resource Scheduling Challenge.} The fifth challenge is to schedule the resources (e.g., CPU threads and memory) for OLTP and OLAP instances effectively for high throughput and data freshness. On the one hand, assigning more resources to OLAP workloads favors high query throughput but may block the OLTP threads due to the limited bandwidth. On the other hand, scheduling more resources to OLTP workloads can accelerate transaction processing but may lead to low data freshness. Hence, the HTAP resource scheduler must balance the trade-off between performance isolation and data freshness as well.



In this work, we provide a comprehensive survey of HTAP databases in three aspects, including HTAP architectures, techniques, and benchmarks. Figure~\ref{fig:overview} presents an overview of HTAP-related techniques. It gives a taxonomy of HTAP architectures and HTAP techniques, respectively. It also presents four components of HTAP benchmarks. We pay particular attention to how existing approaches address the above-mentioned challenges, respectively.





\subsection{HTAP architectures}
We mainly study HTAP databases that utilize row stores and column stores together to handle the mixed workloads of OLTP and OLAP efficiently in a single database system. Based on the storage strategies and processing paradigm, we divide their architectures into four categories as follows:


\noindent \textbf{(1) Primary Row Store+In-Memory Column Store.}  This category of HTAP databases leverages a primary row store as the basis for OLTP workloads and processes OLAP workloads with an in-memory column store. Updates are appended to the delta store, which will be periodically merged with the column store. We review four representatives: Oracle ~\cite{Oracle2015}, SQL Server~\cite{SQLServer2015}, and DB2 BLU~\cite{DB2BLU2013}.







\noindent \textbf{(2) Distributed Row Store+Column Store Replica.} This category relies on a distributed architecture to support HTAP. The master node handles the read-write transactions and asynchronously replicates the logs to the slave nodes. The primary storage relies on a distributed row store, and some slave nodes will be chosen as column-store servers for query acceleration. We introduce two representatives: TiDB~\cite{TiDB2020} and F1 Lightning~ \cite{F12020}.






\noindent \textbf{(3) Primary Row Store+Distributed In-Memory Column Store.} This type of database utilizes a primary with a distributed in-memory column store (IMCS) to enable HTAP. We present a representative: MySQL Heatwave \cite{MySQLHeatWave2021}.








\noindent \textbf{(4) Primary Column Store+Delta Row Store.} This category of databases utilizes an in-memory column store as the basis for OLAP, and handles OLTP with a delta row store, which will be eventually merged into the column store. We introduce SAP HANA~\cite{SAPHANA2012} and Hyper~\cite{Hyper_column}.

\noindent \textit{Other HTAP architectures.} We also review other types of HTAP architectures that complement the major architectures, including (i) row-only HTAP architectures; (ii) column-only HTAP architectures; (iii) Spark-based HTAP architectures; and (iv) cloud-native HTAP architectures.





\subsection{HTAP techniques}


This section takes a deep dive into the key techniques of HTAP databases, including hybrid workload processing, data organization, data synchronization, query optimization, and resource scheduling.


\noindent \textbf{(1) Hybrid Workload Processing.} There are three kinds of hybrid workload processing techniques, including (i) Multi-Version Concurrency Control (MVCC) techniques~\cite{ERMIA_MVCC, Diva_MVCC} (ii) Copy-on-Write (CoW) techniques \cite{Hyper2011, H2TAP2017}; and (iii) Dual-store based processing \cite{Oracle2015, SQLServer2015, MySQLHeatWave2021, SAPHANA2012, TiDB2020}. First, MVCC-based techniques process the hybrid workloads on the same copy of multi-versioned data. Second, the CoW technique relies on snapshotting to support HTAP, where the main process handles the transactions and the forked processes perform the queries. Third, dual-store-based processing utilizes a transactional store for OLTP workloads, and employs an analytical store for OLAP workloads. 

\noindent \textbf{(2) Data Organization.} We introduce two types of data organization techniques, including (i) primary row store with the selected column store and (ii) adaptive hybrid data storage. Both methods adaptively generate the data storage based on the given workloads. The former organizes the data by selectively replicating the data from the row store to the column store, including frequency-based heatmap~\cite{Oracle21c} and integer programming~\cite{hyrise_columnselection}. The latter organizes each table in a hybrid row and columnar format by vertically and horizontally partitioning the tables, including the cost-based approaches~\cite{H2O2014, Casper2019}, a clustering approach~\cite{Peloton2016}, and a deep-learning-based approach~\cite{sigmod/proteus}.

\noindent \textbf{(3) Data Synchronization (DS)} There are two types of DS techniques for various HTAP databases including (i) in-memory delta merge \cite{Oracle2015, SQLServer2015, MySQLHeatWave2021, DB2BLU2013, SAPHANA2012} and (ii) log-based delta merge \cite{IDAA_VLDB2020, TiDB2020, F12020}; The first category periodically merges the newly-inserted in-memory delta data to the memory-based column store.  The second category \cite{TiDB2020, F12020} merges the deltas to the column store based on multi-level delta merging, including log shipping and replaying.



\noindent \textbf{(4) Query Optimization.} We introduce three types of query optimization techniques, including (i) hybrid row/column scan \cite{TiDB2020, SQLServer2015}; (ii) HTAP indexing~\cite{MVPBT2020, P-tree2019} and (iii) CPU/GPU acceleration for HTAP \cite{H2TAP2017, RateupDB2021}. The first type \cite{MySQLHeatWave2021, Oracle21c} optimizes query plans by selecting the access path of a row store or a column store. The second type relies on path-copying and multi-version indexing techniques to speed up HTAP. The third type leverages heterogeneous CPU/GPU architecture to accelerate HTAP workloads.




\noindent \textbf{(5) Resource Scheduling.} There are two types of scheduling techniques: the workload-driven approaches \cite{SAPHANA2012, performance_isolation_2021} and the freshness-driven approach \cite{ResourceScheduling2020}. The former  adjusts the parallelism of OLTP and OLAP threads based on the performance of executed workloads. The latter  \cite{ResourceScheduling2020} switches the execution modes on resource allocation and data exchange for OLTP and OLAP workloads by considering the data freshness.




\subsection{HTAP benchmarks}
We present state-of-the-art end-to-end benchmarks and micro-benchmarks for evaluating HTAP databases. We introduce four end-to-end HTAP benchmarks, including CH-benchmark \cite{CH-benchmark}, HTAPBench \cite{htapbench}, OLxPBench~\cite{conf/icde/OLxPBench}, HATtrick~\cite{sigmod22/HAPtrick}, and HyBench~\cite{pvldb/HyBench}. We focus on the key components, including data generation, HTAP workload, execution rule, and metrics. In addition to the end-to-end benchmarks, we will introduce three synthetic micro-benchmarks for data organization~\cite{Peloton2016, Casper2019, kang2023benchmarking}.

\subsection{Contributions} \label{sec: contribution}
\textbf{Differences with existing surveys.} In this paper, we focus on fundamental techniques of HTAP databases~\cite{sigmod22/HTAP-Database-Survey}. We also summarize the pros and cons of various architectures and techniques. {\"O}zcan  et al.~\cite{HTAPTutorial2019} discussed various HTAP systems rather than the key techniques of HTAP databases. Hieber et al.~\cite{HTAP-Review-2020, htap_journal} reviewed HTAP systems from several dimensions, including architecture, query handling, and concurrency control. However, it lacked an in-depth analysis of HTAP databases and neglected many fundamental HTAP techniques, such as hybrid workload processing, data organization, data synchronization, and query optimization. Compared with a previous survey~\cite{Chinese_HTAPsurvey}, this work has a significant amount of new content: (1) it gives a systematic overview and introduces five HTAP challenges in Section~1; (2) it reviews the evolution of HTAP databases in the history and introduces a trade-off between data freshness and performance isolation in Section 2; (3) it gives a more detailed analysis on the architectures in Section 3, including the cloud-native HTAP architectures; (4) it introduces a new taxonomy of the key techniques and presents each type of techniques in more details in Section 4, including the hybrid workload processing; (5) it presents eight state-of-the-art HTAP benchmarks in Section 5, including OLxPBench~\cite{conf/icde/OLxPBench}, HATtrick~\cite{sigmod22/HAPtrick}, HyBench~\cite{pvldb/HyBench}, and mOLxPBench~\cite{kang2023benchmarking}; (6) it presents six research directions with many new open problems in Section 6, including HTAP for multi-model data analytic~\cite{gart}, serving atop HTAP~\cite{HSTAP, kang2023nhtapdb}, and cloud-native HTAP techniques~\cite{AlloyDB2022, SingleStore2021, Snowflake2022}.





To summarize, we make the following contributions:

\begin{enumerate}[leftmargin=*]
\setlength\itemsep{0em}
    \item We survey HTAP databases. We introduce a taxonomy of state-of-the-art HTAP databases according to their storage architectures. We also discuss their pros and cons on performance, scalability, and data freshness. 
    
    \item We summarize HTAP techniques. We take a deep dive into the key HTAP techniques concerning hybrid workload processing, data organization, data synchronization, query optimization, and resource scheduling.
    
    \item We review HTAP benchmarks. We introduce the state-of-the-art benchmarks on HTAP databases. We present their schema, workloads, execution rules, and metrics.
    
    \item We provide new research challenges and discuss future directions, including data organization for distributed HTAP databases, HTAP query optimization, and cloud-native HTAP techniques.
\end{enumerate}
\section{Background of HTAP Databases}
In this section, we introduce the background of HTAP databases. We first review the evolution of HTAP databases by introducing four phases of HTAP development and applications, and then we introduce a trade-off between data freshness and performance isolation. 
\begin{figure*}[!t]
	\centering
	\includegraphics[width=1.0\linewidth]{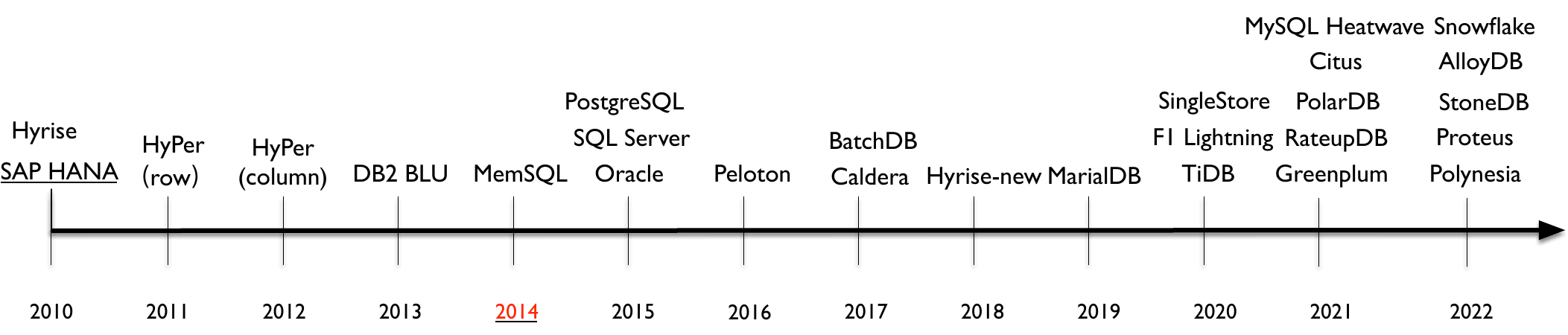}
	\vspace{-1.5em}
 	\caption{A Timeline of HTAP databases that first released the HTAP functionality in the literature}	\label{fig:history}
	\vspace{-1em}
\end{figure*}

\subsection{The Evolution of HTAP Databases}

Figure~\ref{fig:history} depicts a timeline of the HTAP databases between 2010 and 2022. We place the systems in the year when they first released the HTAP functionality from the literature or from the publicly released material. By investigating these HTAP databases, we mainly classify the development of HTAP databases into four phases as follows:

\textit{Year 2010-2014: Standalone Column-based HTAP databases.} In the first phase, HTAP databases mainly adopted standalone column-based databases, representatives are SAP HANA~\cite{saphana_2012}, Hyrise~\cite{Hyrise2010}, Hyper~\cite{Hyper2011}. Since HTAP was not formally defined at that time, they named such a technique as "hybrid OLTP\&OLAP"~\cite{Hyper2011} or "OLxP"~\cite{PostgreSQL2015}. Back then, the applications they targeted were mainly analytical applications with read-heavy transactional workloads, such as the ERP applications~\cite{saphana_2012}.

\textit{Year 2014-2019: Standalone Row-based HTAP databases.} The year 2014 is the time when the HTAP term was formally defined by a Gartner report \cite{Gartner2014htap}. It initially defined HTAP as an application architecture that utilized in-memory computing technologies to enable hybrid processing on the same in-memory data store. In 2018, Gartner extended the HTAP concept to "In-Process HTAP"~\cite{gartner2018}, which supported weaving hybrid workload processing techniques together as needed to accomplish the business task. Such a new definition indicated that HTAP is no longer limited to in-memory computing techniques, which significantly expanded the HTAP applications. During this period, major relational databases extended the primary row store with a column store. To name a few, DB2 BLU~\cite{DB2BLU2013}, MemSQL~\cite{MemSQL2014}, SQL Server~\cite{SQLServer2015}, Oracle~\cite{Oracle2015}, PostgreSQL~\cite{PostgreSQL2015} and MarialDB~\cite{MariaDB2021}. The applications they targeted were medium-scale transaction processing applications with real-time data analytics, such as banking and finance services~\cite{Banking_HTAP} and applications of fraud detection~\cite{FraudDetection2, FraudDetection1}.

\textit{Year 2019-2022: Distributed HTAP databases.} In the third phase, there emerged many distributed HTAP databases, such as SingleStore~\cite{SingleStore2021}, Citus~\cite{cubukcu2021citus}, F1 Lightning~\cite{F12020}, and TiDB~\cite{TiDB2020}. On the one hand, these databases embraced the NewSQL movement~\cite{NewSQL} by developing distributed SQL-based transaction processing systems with high scalability and strong consistency. On the other hand, they caught up with the HTAP wave by expanding their capacity of scalability and consistency to HTAP, e.g., adding distributed columnar storage and unified data replication. Generally, they are suitable for large-scale data-intensive applications such as E-commerce~\cite{OceanBase}, Internet of Things (IoT)~\cite{IoT}, and real-time social media ~\cite{SocialMedia}.

\textit{Year 2022-present: Cloud-Native HTAP databases.} As cloud databases are proliferating, there have been emerging many cloud-native HTAP databases~\cite{CloudDatabases}. Two representatives are AlloyDB~\cite{AlloyDB2022} and Snowflake~\cite{Snowflake2022}. With the disaggregation of computing and storage, they enable HTAP with high elasticity, high availability, and multi-tenancy. First, since compute and storage resources can be scheduled on demand individually, they provide high elasticity for HTAP. Second, as the data is replicated across many availability zones and is backed by the scalable cloud service, cluster and node failures can be recovered quickly. Third, as the resources are virtualized and shared by multiple tenants, they are more cost-efficient. Since Gartner predicted~\cite{Gartner2021Cloud} that the revenue from the cloud databases will account for 50\% of total DBMS market revenue in the near future, we believe cloud-native HTAP databases will find a wide spectrum of applications.

\begin{figure}[!t]
	\centering
	\includegraphics[width=1.0\linewidth]{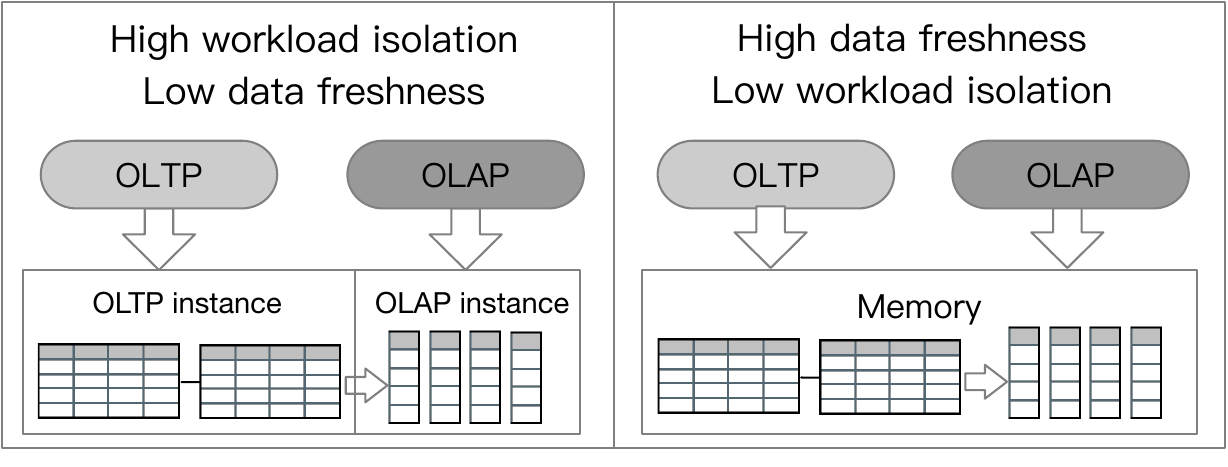}
	\vspace{-1.5em}
 	\caption{A trade-off between data freshness and performance isolation}\label{fig:tradeoff}
	\vspace{-1em}
\end{figure}


\subsection{A Trade-off between Freshness and Isolation} \label{subsec:trade-off}
\textbf{Data freshness.} As OLTP workloads are updating the data, HTAP databases need to guarantee that the fresh data is accessed by the analytical queries. Hence, \textit{the data freshness refers to the lag time between the analytics and transactions}. One way of quantifying data freshness is to take the maximum value of the timestamp differences between the result sets of the OLAP client and OLTP client~\cite{pvldb/HyBench}. Additionally, the data replication latency can also reflect the data freshness. Particularly, with the separated transactional store and analytical store, the fresh data is periodically replicated in the analytical store. Since many approaches~\cite{TiDB2020, Vegito2021, F12020} must merge the newly-updated data to the analytical store before performing the data analytics, it measures how fast the recently committed transactions are synchronizing to the analytical store such that the analyzed data is fresh.

\noindent \textbf{Performance isolation.} In HTAP databases, \textit{performance isolation~\cite{htapbench, CH-benchmark, TiDB2020, TPCTC2014, Vegito2021} refers to the system's capacity of reducing the performance degradation when processing the dynamic hybrid workload}. In other words, it reflects how well the systems can reduce the interference between OLTP and OLAP workloads when executing them concurrently. Regardless of the isolation architectures (in the same node or different nodes), such metrics directly reflect HTAP databases' performance of handling hybrid workloads w.r.t. the same number of OLTP/OLAP clients and the same scale factor of a dataset.



\noindent \textbf{A Trade-off.} HTAP databases must balance a trade-off between data freshness and performance isolation. As shown in Figure~\ref{fig:tradeoff}, handling the hybrid workloads with separated instances can provide high-performance isolation but may lead to low data freshness, hence the outdated results of data analytics. For instance, TiDB~\cite{TiDB2020} only degraded up to 10\% of the performance while it took up to 1000 ms to apply the change logs for data replication. Handling the hybrid workloads in a unified memory space of a single sever favors high data freshness but it can greatly degrade the performance. For instance, Hyper~\cite{Hyper_column} took only microseconds to create a fresh snapshot, but it degraded up to 40\% of the performance. To quantify the effectiveness of performance isolation, the degradation percentage~\cite{TiDB2020, TPCTC2014} of transactional/analytical throughput is mostly used. One way of quantifying the performance isolation is to compare the performance between a sequential execution and a hybrid execution with the hybrid workload\cite{pvldb/HyBench}. The lower the metric is, the better the performance isolation is.

\section{HTAP Architectures}
\label{sec:HTAP-architecture}
\begin{figure*}[!t]
	\centering
	\includegraphics[width=1.0\linewidth]{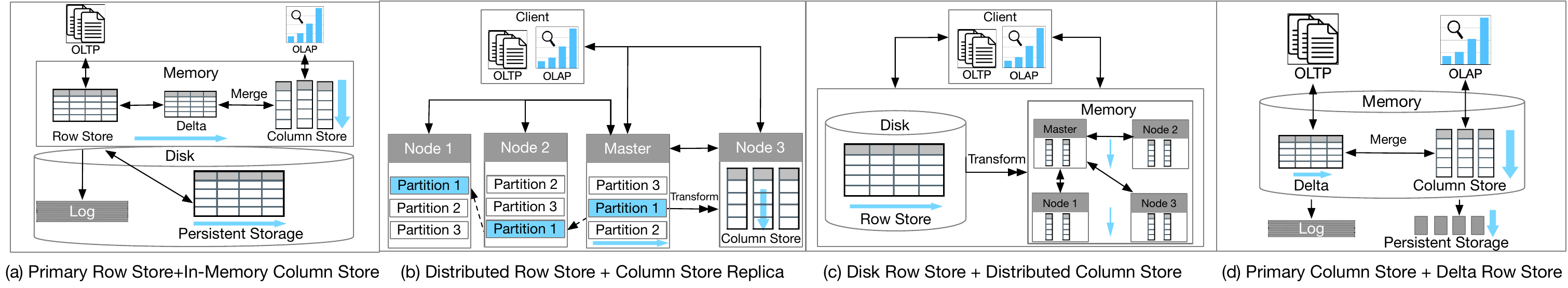}
 	\caption{A Taxonomy of State-Of-The-Art HTAP Databases based on the Storage Architecture and Processing Paradigm}	\label{fig:architecture}
\end{figure*}

\begin{table*}[!t]
\footnotesize
\renewcommand\arraystretch{1.2}
\caption{A Classification of State-Of-The-Art HTAP Databases based on the Storage Architecture \label{table:classification}} 
\begin{tabular}{|p{15 em}|C {11.5 em}|C {11 em}|C {11.5 em}|C {8 em}|}
\hline
Category & HTAP databases & OLTP Paradigm & OLAP Paradigm & Delta Store\\ \hline
Primary Row Store + In-Memory Column Store  & \begin{tabular}[c]{@{}c@{}}Oracle~\cite{Oracle2015}, \\  SQL Server~\cite{SQLServer2015}  \end{tabular} & Standalone Row-wise MVCC         & Standalone Columnar Scan with Delta Traverse & In-Memory Table                                                              \\ \hline
Distributed Row Store + Column Store Replica      & \begin{tabular}[c]{@{}c@{}}TiDB~\cite{TiDB2020} \\F1 Lightning~\cite{F12020} \end{tabular}                                                                 & Distributed Row-wise 2PC + Paxos              &    Distributed Columnar Scan with Log Replay            & B-tree, Change Log                                                                 \\ \hline
Primary Row Store + Distributed In-Memory Column Store & MySQL Heatwave\cite{MySQLHeatWave2021}                                                                     & Standalone Row-wise MVCC           & Distributed Columnar Scan with Log Replay           & Change Log                                                                  \\ \hline
Primary Column Store + Delta Row Store & SAP HANA~\cite{SAPHANA2012}, Hyper~\cite{Hyper_column}                                                                           & Cache-based Row-wise MVCC          & In-Memory Columnar Scan with Delta Traverse              & In-Memory Dictionary                                                               \\ \hline
\end{tabular}
\end{table*}

In this work, we focus on the HTAP architectures with a row store and a column store in a single DBMS. We do not consider the loosely-coupled HTAP architectures~\cite{ HTAPTutorial2019} as those architectures employ multiple databases to form an HTAP solution that entails a costly ETL process. We classify the tightly-coupled HTAP architectures into four categories: (a) Primary Row Store and In-Memory Column Store; (b) Distributed Row Store and Column Store Replica; (c) Primary Row Store and Distributed In-Memory Column Store; and (d) Primary Column Store and Delta Row Store. Table \ref{table:classification} presents their representative databases, OLTP paradigm, OLAP paradigm, and delta storage.

\subsection{Primary Row Store+In-Memory Column Store} \label{subsec: arc-a}
This category of databases~\cite{Oracle2015, SQLServer2015, DB2BLU2013} leverages a primary row store and an in-memory column store in a single node. It relies on a row store for handling the OLTP workload, and the data will be persisted to the disk in a row-wise page store. Column store is optimized for OLAP workload with compression techniques~\cite{ColumnDBMSs, Hyrise_encoding} and standalone columnar scans~\cite{Oracle2015, SQLServer2015, DB2BLU2013}. An in-memory delta store is utilized to record the recent DML operations, which will be merged into the column store periodically. For transaction processing, read/write transactions will be handled by the row engine with the ACID guarantee based on MVCC; data updates will also be recorded in the delta store. For analytical processing, long-running queries are processed using the columnar scan, and the delta data that has not been merged should be traversed to access the fresh data. This category of system has high analytical throughput because the OLAP workloads are processed using an in-memory column store. Data freshness is also high because the column store engine can access the latest data in the main memory. However, limited by memory capacity, the scalability of this category of system is low. In addition, because transactional and analytical workloads are processed in the same machine, the isolation of the system is low. 


\subsubsection{Representatives of HTAP Databases}
With architecture (a), Oracle in-memory dual-format database \cite{Oracle2015, Oracle21c} combines the row-based buffer cache with an in-memory column store (IMCS) to handle OLTP and OLAP workloads simultaneously. IMCS consists of in-memory compression units (IMCUs), and each IMCU is immutable and can only be populated from the buffer; the data changes are cached in the snapshot metadata unit (SMU), and each IMCU is associated with one SMU. To merge the updates to the IMCU, it must create a new IMCU that incorporates the data in the corresponding SMU. Another representative is SQL Server \cite{SQLServer2015}, where an in-memory row-based engine, called Hekaton~\cite{hekaton2013}, is integrated for handling OLTP workloads. Its underlying storage is based on a disk-based row store for logging the operations and persisting the data. It also builds the column store index (CSI) for handling complex queries. Different from Oracle, the in-memory columns in CSI that are infrequently accessed will be compressed and persisted to the disk, and CSI is updatable. Besides, the data changes are appended to the tail of the in-memory table, and they are indexed by a tail index, e.g., a B-tree. By scanning the tail index, data changes will be periodically merged into the column store. DB2 also adopts such an architecture; it has been deeply integrated with an IMCS accelerator called DB2 BLU~\cite{DB2BLU2013} to incorporate advanced column-based techniques such as compression-based operations and single-instruction multiple-data (SIMD) instructions. BLU also supports updating the column store. 





\subsubsection{Challenges \& Opportunities}
HTAP databases with architecture (a) need to improve the scalability and performance isolation while maintaining high performance and high data freshness. The main challenge is how to scale and isolate the OLAP workloads using the column store. One possible solution is to scale the OLAP workloads with a distributed in-memory architecture~\cite{StoneDB2022}. In addition, placing which columns into the column store with a memory budget is also challenging as it is an NP-hard problem~\cite{hyrise_columnselection}.


\subsection{Distributed Row Store+Column Store Replica}
The second architecture is a distributed cluster including a master node and multiple secondary nodes. The master node handles the read/write transactions; the secondary nodes are read-only. Particularly, when handling the transaction requests, the master node asynchronously replicates the logs to the secondary nodes for data synchronization. The primary node adopts a row store, certain secondary nodes will be chosen to adopt a column store for query acceleration. Transactions are handled in a distributed way for high scalability; complex queries are performed in the secondary nodes with a column store. For the pros, it has high performance isolation and high scalability as the mixed workloads are processed on separated nodes. For the cons, the data freshness is low since newly-updated data may not have been merged into the column store.

\subsubsection{Representatives of HTAP Databases}
Two representatives with architecture (a) are TiDB~\cite{TiDB2020} and F1 Lightning~\cite{F12020}. TiDB is a Raft-based distributed HTAP database~\cite{Raft}, which consists of multiple row-based nodes, called TiKV nodes, among which a leader node handles the read-write transactions. The leader node asynchronously sends the data to other follower nodes; the follower nodes only serve the read transactions and consistently communicate with the leader. Only if the leader node has failures,  a follower node can become a leader by voting. For the data storage, the underlying data is partitioned into multiple row-based regions. One or more servers will be selected as learner nodes that store columnar replicas for analytical processing. Learner nodes are read-only and do not participate in voting. It adopts a global 2-Phase-Commit (2PC) protocol~\cite{Pecolator} to handle distributed transactions based on a global time stamp. For analytical processing, it develops a centralized cost-based optimizer that supports cross-engine query processing where queries can be pushed down to either a row engine or a column engine. TiDB relies on the Raft protocol for data replication~\cite{Raft}. The master node asynchronously replicates logs to the follower and learner nodes for log replaying, and it builds a delta merge tree to track the changes and merges them to the column store periodically. F1 Lightning~\cite{F12020} is another representative, which integrates a data replication service into an OLTP engine that is built on top of Spanner~\cite{Spanner_OSDI_2012}, a distributed OLTP database with strong consistency and high scalability that organizes the row-based partitions in multiple regions. The lightning server contains a component, called changepump, which uses the change data capture mechanism to detect new changes, then transforms them from row-wise format to columnar format and merges them into the storage. The memory-resident deltas are row-wise B-tree, and Lightning periodically checkpoints memory-resident deltas to disk. When the deltas are too large, Lightning merges and transforms them into the column store. The log-structured merge (LSM) reader merges both memory-resident deltas and disk-based deltas by merging and collapsing. Particularly, merging deduplicates changes in the deltas and copies distinct versions to the new delta; collapsing combines multiple versions of the same key into a single version. F1 lightning adopts a Paxos-based 2PC protocol to handle distributed transactions; a distributed query engine~\cite{Paxos}, F1 Query~\cite{F1-Query-VLDB2018}, is employed for query processing. For the specified timestamp that is in the query window (normally 10 hours), the pushdown evaluator reads the corresponding data from the columnar file in the lightning server, and then obtains other data from Spanner.

 

\subsubsection{Challenges \& Opportunities}
HTAP databases with architecture (b) need to increase the data freshness due to the high overhead of log shipping, transformation, and delta merging in the distributed architecture. The main challenge is how to efficiently merge the delta files to the column store. Two possible solutions are to (1) develop a memory-based delta logging and shipping~\cite{Vegito2021} and (2) design new indexing techniques for delta merging~\cite{Laser2021, Umzi_LSM_index}. Moreover, how to effectively organize the data (e.g., data layout, data placement, and column compression) for a distributed HTAP database is also a challenging task~\cite{sigmod/proteus, Tiresias}.

\subsection{Primary Row Store+Distributed IMCS}
This kind of HTAP databases~\cite{MySQLHeatWave2021, StoneDB2022} utilizes a row store with a distributed in-memory column-store (IMCS) to support HTAP. Two main differences between architecture (a) and (c) are about the OLAP paradigm and delta store. For the OLAP paradigm, the former one adopts the standalone in-memory columnar scan with delta traverse while the latter one relies on distributed in-memory columnar scan with log replay. Regarding the delta store, architecture (a) employs the in-memory table while architecture (c) uses the change log, so they have the different data synchronization methods (See Section \ref{sec:DS}).


With architecture (c), the row engine processes the OLTP workloads, and the IMCS handles the query processing. The columnar data is extracted from the row store, and the hot data resides in IMCS and the cold data will be evicted to disk. For the pros, it has high performance isolation as the hybrid workloads are processed in different nodes. Moreover, it has a high OLAP throughput and scalability because of the distributed IMCS. For the cons, it has medium or low data freshness, depending on the deployment mode of the IMCS cluster (e.g., on-premise or on-cloud). In addition, it has low horizontal scalability on OLTP due to a standalone row store for transaction processing.  



\subsubsection{Representatives of HTAP Databases}
MySQL Heatwave \cite{MySQLHeatWave2021} is a representative that employs architecture (c). Specifically, it tightly couples the MySQL database with a distributed IMCS cluster, called Heatwave, to enable real-time analytics. Transactions are fully executed in the MySQL database. Columns that are frequently accessed will be loaded into the Heatwave. When a complex query comes in, it can be pushed down to Heatwave for query acceleration. The columnar data is extracted from the MySQL database, and the hot data resides in the Heatwave cluster. For data synchronization, the latest transaction data will be automatically transformed and merged to the column store in three cases: (i) every 200 milliseconds; (ii) when the buffer size reaches 64 MB, or (iii) when the queries need to access the latest updated data. The heatwave cluster also developed the auto-pilot service to automate the processes of data partition, query execution and scheduling. 



\subsubsection{Challenges \& Opportunities}
HTAP databases with architecture (c) need to increase the data freshness due to the distributed in-memory column store. The main challenge is how to balance the data freshness and OLAP throughput. Possible solutions are to design (1) cost functions or (2) ML models for column data management, including column selection and compression~\cite{CodecDB}.

\subsection{Primary Column Store+Delta Row Store} \label{subsec:arc-d}
This category of databases utilizes the primary column store as the basis for OLAP, and handles OLTP with a delta row store. The primary column store stores the whole data in the main memory. Data updates are appended to the row-based delta store. The system periodically merges the delta data into the column store. For the pros, it has high data freshness as the hybrid workloads are processed in the main memory. Moreover, it has a high OLAP throughput because of the primary column store. For the cons, it has low OLTP scalability due to the delta-based row store. In addition, it has low-performance isolation due to the standalone architecture for hybrid workload processing.

\subsubsection{Representatives of HTAP Databases}
SAP HANA~\cite{saphana_2012, SAPHANA2012} and Hyper~\cite{Hyper2011, Hyper_column} are two representatives of architecture (d). SAP HANA divides the in-memory data store into three layers: L1-delta, L2-delta, and Main. The L1-delta keeps data updates in a row-wise format. When a threshold is reached (e.g., 100k tuples), the data in L1-delta is transformed and merged to L2-cache based on a local dictionary. The L2-delta transforms the data into columnar data, then merges the data into the main column store based on a global dictionary. Finally, the columnar data is persisted in the disk storage. For high data freshness, the OLAP client will traverse the delta records when scanning the column store. In addition, for non-transformed records, it can use in-place updates for the delta store. To update transformed records, it replaces an update operation with a delete and an insert operations on the delta store. Hyper~\cite{Hyper2011, Hyper_column} was initially based on a row store. Now it has supported the architecture of a primary column store with a delta row store. Specifically, it uses the buffer to handle concurrent transactions based on an MVCC protocol. The version vector stores all the versions for each unique tuple. Instead of periodically merging the data in the version vector to the column store, Hyper adopts transaction-level garbage collection, where all of the versions that were generated by the transactions can be safely removed after the transactions have been applied to the column store. Finally, the columnar data is updated in-place by applying all the committed transactions.


\subsubsection{Challenges \& Opportunities}
For HTAP databases with architecture (d), two main problems need to be addressed. (1) they need to increase the OLTP scalability due to the delta-based transaction processing. (2) they need to increase performance isolation due to the unified memory pool for HTAP. The main challenge is how to traverse the delta storage efficiently while keeping high throughput for HTAP. Although they have offered certain scaling-out solutions~\cite{saphana2015, ScyPer}, these approaches need to be further justified due to the centralized transaction scheduler.  


\subsection{Applications of HTAP Architectures} HTAP databases with the above-mentioned architectures have their merits and demerits. Hence, ``one HTAP database cannot fit all", especially for different applications. By comparing their pros and cons, we summarize the gained insights and give the recommendations as follows.

(1) Architecture (a) is suitable for the applications that require high throughput and data freshness, but the demand for scalability is not high. Representatives are the banking and finance services~\cite{Banking_HTAP} that need to process and analyze the customers' transactions efficiently. Therefore, these applications have a high requirement on the system throughput. Since the number of target customers is almost fixed, these applications do not require high scalability.


(2) Architecture (b) is ideal for applications that require high scalability and can have tolerable data freshness. A representative is an E-commerce application with real-time data analytics. Such applications need to process a large number of concurrent transactions from multiple customers, especially for holidays, e.g., Double-Eleven in Alibaba~\cite{OceanBase}. Therefore, the scalability requirement must be fulfilled. Nevertheless, such applications do not force zero freshness as it is acceptable that there is any inconsistency in a short period of time, e.g., customer retention rate.


(3) Architecture (c) is suitable for applications that require high analytical throughput and scalability, but the demand for data freshness is not high. A representative is the IoT applications with real-time data analytics~\cite{IoT}. On the one hand, these applications have a high requirement for analytical throughput and scalability. Thus, a distributed in-memory column store is a good fit for such cases. On the other hand, the data updates in these scenarios are rare, thus a standalone row store is sufficient.


(4) Architecture (d) is a good fit for applications that require high analytical throughput and high data freshness. A representative is real-time fraud detection~\cite{FraudDetection1, FraudDetection2}, which requires high data freshness as they can not tolerate any fraud, which could cause significant consequences. Nevertheless, the scalability requirement is not high.




\subsection{Other Architectures}
Other than the four main types of HTAP architectures, there are other HTAP architectures, including (1) row-only HTAP architectures; (2) column-only HTAP architectures; (3) Spark-based HTAP architectures; and (4) cloud-native HTAP architectures. We summarize them as follows:



(1) Row-only HTAP architectures~\cite{Hyper2011, BatchDB2017} rely on the purely row store to enable HTAP. For instance, the first version of Hyper~\cite{Hyper2011} employed a copy-on-write mechanism to fork an OLAP process to operate on a separate snapshot. Its main process handles the transactions simultaneously, and both the OLTP and OLAP processes are based on the row-based data store. BatchDB~\cite{BatchDB2017} leverages the primary-secondary replication, where the primary row-wise replica is used for OLTP, and the secondary row-wise replica is in charge of OLAP. It uses a batch-based propagation mechanism to synchronize the updates from the primary replica to the secondary replica. This type of architecture relies heavily on the row-based query processing in the main memory to process the hybrid workload. Thus, both the freshness and OLTP throughput are high. However, the major drawback is that the analytical throughput is much lower compared to the column-based query processing.

(2) Column-only HTAP architectures~\cite{H2TAP2017, Hyrise2010, RateupDB2021, L-store2016} rely solely on a column store to support HTAP. For example, Hyrise~\cite{Hyrise2010} initially employed an adaptive columnar layout to support HTAP, and the basic idea is to use narrower column groups to handle OLAP workloads and leverage wider column groups to handle OLTP workloads. The latest version of Hyrise~\cite{Hyrise2019} develops the chunk-based column store, a PAX~\cite{PAX}-like data layout that horizontally divides a table into partitions (a.k.a., row groups), and each partition is organized in columns. NoisePage~\cite{NoisePage} (previously named Peloton~\cite{Peloton2016}), a self-driving columnar database~\cite{NoisePage_TrainingData, NoisePage_Workload, NoisePage_MB2, NoisePage_SelfDriving}, also adopts such a data layout based on Apache Arrow~\cite{Arrow} format. Caldera~\cite{H2TAP2017} relies on the copy-on-write mechanism with CPU/GPU architecture where OLTP workloads are handled by multi-threads of CPU and OLAP workloads are executed with GPU in parallel. Another case is RateupDB~\cite{RateupDB2021}, which also adopts the CPU/GPU architecture.  Different from Caldera~\cite{H2TAP2017}, it relies on the primary-secondary replication to enable HTAP. TiQue is a recent work that adds the transaction logic in SQL and enables HTAP on top of a column store. The basic idea is to add the delta table and transaction metadata for an analytical database, thereby accelerating the transaction processing. Overall, column-only HTAP architectures have a high analytical throughput, but the data updates and data synchronization can significantly affect the system's performance.



(3) Spark-based HTAP architectures. Such systems~\cite{barber2017evolving,mozafari2017snappydata} support HTAP by combining an OLTP engine and an OLAP engine. The OLTP engine is dominated by columnar databases such as HBase~\cite{team2016apache}, and the OLAP engine mainly uses a Spark~\cite{zaharia2010spark} engine. Both engines share the data in a distributed file system (such as HDFS~\cite{borthakur2008hdfs}). For example, Splice Machine~\cite{SpliceMachine} and Phoenix~\cite{Phoenix} handle data updates based on HBase, and leverage Spark for big data analytics. SnappyData~\cite{mozafari2017snappydata} integrates a transactional engine, Apache Geode~\cite{Geode}, into the Spark for processing streaming, transactional, and interactive analysis simultaneously. Wildfire \cite{barber2017evolving} tightly couples the Spark with a distributed transaction engine. In the upper layer, it provides a unified interface to take as input the hybrid workloads. The middle layer schedules the OLTP workloads with Scala API to the transactional engines, and routes the Spark SQL queries to Spark executors. In the storage layer, they organize the data in the LSM-tree format with an indexing support~\cite{Umzi_LSM_index}. The Spark-based HTAP systems have high scalability and are suitable for big data analytics with modest data updates. However, the data freshness is low due to the shared file storage.


(4) Cloud-Native HTAP architectures~\cite{PolarDB-X-ICDE2022, PolarDB_Serverless,AlloyDB2022, SingleStore2021, Snowflake2022} rely on the disaggregation of compute and storage to enable HTAP. For instance, AlloyDB~\cite{AlloyDB2022} is a PostgreSQL-compatible cloud-native HTAP service. In its compute layer, the compute nodes rely on machine learning to convert the in-memory data from the row format to columnar format for query acceleration. In its storage layer, AlloyDB has developed an elastic log storage service that organizes the data in shards and asynchronously materializes the WAL records to fresh pages, and the dirty pages in the compute layer are never flushed into the storage~\cite{Aurora}. Another example is SingleStore (the successor of MemSQL~\cite{MemSQL2014}), which proposes a unified table storage structure based on a cloud-native architecture. It relies on a distributed in-memory row store for handing updates. The data is persisted to the disk-based columnar storage, and is organized as LSM trees in the storage layer; the secondary hash indexes are built for accelerating point queries. Snowflake~\cite{Snowflake_2016}, which is a cloud-native OLAP database, has utilized its metadata storage to implement an HTAP solution, called Unistore~\cite{Snowflake2022}, supporting real-time data analytics. PolarDB-IMCI builds an in-memory column index in a disagreggation architecture of PolarDB~\cite{PolarDB_Serverless}. Data synchronization is performed by replaying physical redo logs from the shared storage. Cloud-native architectures empower HTAP with high elasticity, availability, and multi-tenancy. However, the data freshness is low due to the philosophy of "log is the database", especially when the log has not been replayed in the storage layer.



\section{HTAP Techniques}
\label{sec:HTAP_Techniques}
\begin{table*}[!t]\vspace{-1.5em}
\footnotesize
\renewcommand\arraystretch{1.15}
\caption{An Overview of HTAP Techniques\label{table:techniques}}
\begin{tabular}{|c|c|c|c |c |}
\hline 
Task Type                               & Main Method                             & Key Technique                                                                                                    &  Pros & Cons \\ \hline
\multirow{3}{*}{\begin{tabular}[c]{@{}c@{}}Hybrid Workload \\ Processing\end{tabular}}    &  MVCC-based HTAP   &  Version Chains for OLTP and OLAP~\cite{ERMIA_MVCC}
                                            & High Freshness      & Long Version Chains    \\ \cline{2-5}    &  Copy-on-Write based HTAP           &  Snapshotting OLTP for OLAP~\cite{Hyper2011}
                                            & High Freshness      & Large Memory Size   
                                                                         
\\ \cline{2-5}   &       Dual-Store based HTAP & Separated Stores for OLTP and OLAP~\cite{Oracle2015}  & High Isolation        & High Sync Cost       
                                        \\ \hline
\multirow{4}{*}{\begin{tabular}[c]{@{}c@{}}Data \\ Organization\end{tabular}} &   

\multirow{2}{*}{\begin{tabular}[c]{@{}c@{}}Primary Row Store with \\ Selected Column Store\end{tabular}} &  Frequency-based Heatmap~\cite{Oracle21c}  &  Arbitrary Queries     & Low Utility  
\\ \cline{3-5} &  & Cost-Based Linear Programming~\cite{hyrise_columnselection} & High Utility    & Single-Table Queries         \\  \cline{2-5}    
                                        &  \multirow{2}{*}{\begin{tabular}[c]{@{}c@{}}Adaptive Hybrid \\ Data Storage\end{tabular}}  & Cost Functions~\cite{sigmod/proteus,H2O2014,Peloton2016,Casper2019} & High Efficiency   &  Low Utility    \\
                    
                    \cline{3-5} &  & Machine Learning Models~\cite{sigmod/proteus, Tiresias}  & High Utility  & Training Overhead             \\  \cline{1-5}    
                                                            
\multirow{5}{*}{\begin{tabular}[c]{@{}c@{}}Data \\ Synchronization\end{tabular}}   & \multirow{3}{*}{In-Memory Delta Merging}                & Threshold-based Merging~\cite{Oracle2015}                                                                  &  Fast Insertion      &  Slow Lookup

                                           \\ \cline{3-5} &  & Delete table based Merging~\cite{SQLServer2015}                                   & Fast Lookup  & Slow Insertion
                                                                                      \\ \cline{3-5} &  &Dictionary-based Merging~\cite{SAPHANA2012}                                                                                                  & High Efficiency  & Low Scalability  
                                                                                      
\\ \cline{2-5} 
                                        & \multirow{2}{*}{Log-based Delta Merging}  & Multi-Level Delta Merging~\cite{TiDB2020,F12020}                                          & High Scalability    & High Merge Cost 
                                           \\ \cline{3-5} &  & Change Data Capture~\cite{IDAA_VLDB2020, MySQLHeatWave2021}                                                                & High Isolation  & High Latency 
                                        \\ \hline
\multirow{4}{*}{\begin{tabular}[c]{@{}c@{}}Query \\ Optimization\end{tabular}}             

& \multirow{2}{*}{Hybrid Row/Column Scan}   & Cost-based Execution~\cite{TiDB2020, hybridindex2018, mentis} & High Utility     & Large Search Space    
\\ \cline{3-5} &  & Rule-based Execution~\cite{Oracle2015, SQLServer2015}   & High Efficiency  & Low Utility     
                                        \\ \cline{2-5}                                       &    \multirow{2}{*}{HTAP Indexing}   &  Parallel Binary Tree~\cite{P-tree2019} & High Throughput  & Large Memory Size
                                       
                                           \\ \cline{3-5} &    & Multi-Version Partitioned B-tree~\cite{MVPBT2020}                                                                                                 & High Scalability  & Low Throughput  
                                                                               \\ \cline{2-5}
                            
                                              & CPU/GPU Acceleration                          & CPU for OLTP, GPU for OLAP \cite{H2TAP2017, RateupDB2021}                                         & High AP Throughput     & Low Freshness       
                                        
                                        \\ \hline
                                        
                 \multirow{2}{*}{\begin{tabular}[c]{@{}c@{}}Resource \\ Scheduling\end{tabular}}             & Freshness-Driven Method      &  Execution Mode Switching~\cite{ResourceScheduling2020}     &            High Freshness             &            Low Throughput     
                 \\ \cline{2-5} 
                                & Workload-Driven Method                            & Rule-Based Resource Assignment~\cite{performance_isolation_2021}                                                                                                       & High Throughput   & Low Freshness     
                 \\ \hline
                 \end{tabular}
\end{table*}

\begin{figure}[!t]
	\centering
	\includegraphics[width=0.8\linewidth]{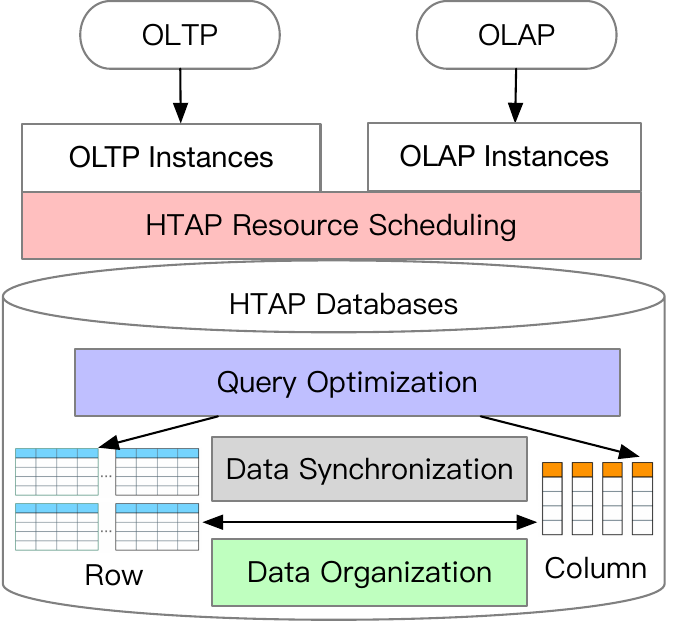}
 	\caption{The Key Techniques of HTAP Databases} \label{fig:HTAP_techniques}
	\vspace{-1em}
\end{figure}

As shown in Table \ref{table:techniques}, we summarize five types of HTAP techniques, including (1) hybrid workload processing; (2) data organization; (3) data synchronization; (4) query optimization; and (5) resource scheduling. It also presents their main methods, key techniques, pros and cons. Figure~\ref{fig:HTAP_techniques} illustrates an overview of the HTAP techniques. From a top-down perspective, (1) OLTP and OLAP workloads are handled by hybrid workload processing. (2) HTAP resource scheduling dynamically assigns the resources to the hybrid workload. (3) Query optimization relies on row and column stores to execute the query. (4) Data synchronization merges the updates from the row store to the column store. (5) Data organization adaptively organizes the hybrid storage.

\subsection{Hybrid Workload Processing}
\begin{figure}[!t]
	\centering
	\includegraphics[width=1.0\linewidth]{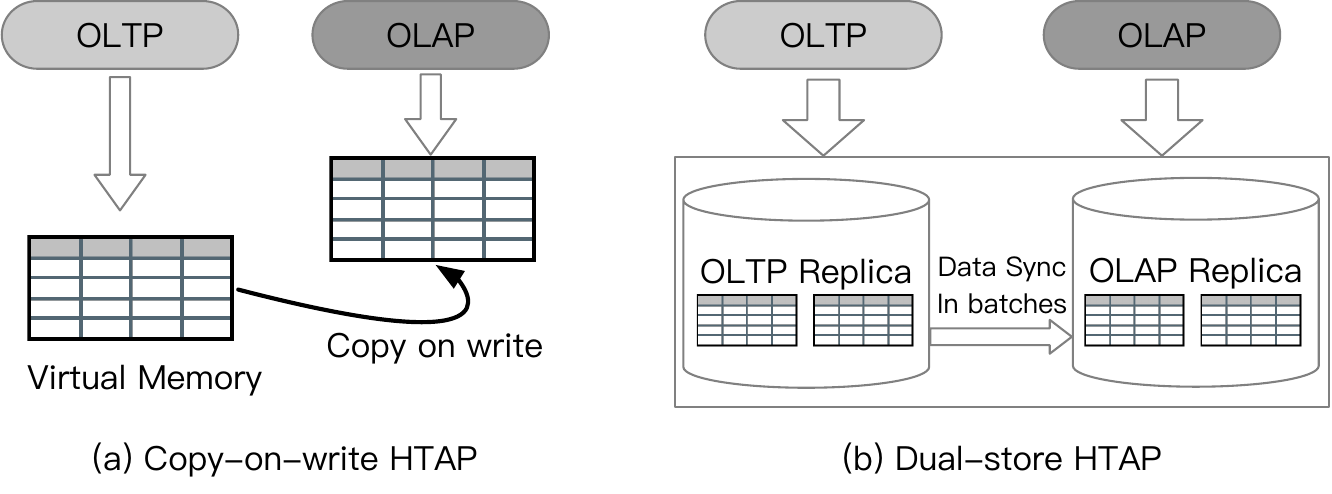}
	\vspace{-1.5em}
 	\caption{Hybrid Processing based on Copy-on-Write and Dual-Store}	\label{fig:HP-CoW}
	\vspace{-1em}
\end{figure}

Hybrid workload processing aims to handle the mixed workload under a unified architecture. It mainly consists of three types of techniques: (1) MVCC-based HTAP; (2) copy-on-write-based HTAP; and (3) dual-store-based HTAP. 

\subsubsection{MVCC-based HTAP} \label{subsec:MVCC_HTAP}

Multi-Version Concurrency Control (MVCC) is the most widely adopted transaction management technique in major relational DBMSs~\cite{empiricalMVCC2017}. However, it is challenging to support HTAP solely based on MVCC. This is because HTAP workloads will lead to frequent version traversing and cleaning of stale data versions. Therefore, the main challenge is how to reduce the resource contention and improve the query performance by efficiently traversing and reclaiming the version chains~\cite{alhomssi2023scalable}. To make the MVCC-based databases more HTAP friendly, Weaver~\cite{Weaver_MVCC} utilizes the frag skip lists to speed up the version chain lookup by using two pointers for each node, which can accelerate the in-chain version traversing and cross-chain version traversing, respectively; Diva~\cite{Diva_MVCC} separates the version searching and version cleaning by maintaining a provisional index and performing an interval-based version clearning separately, which results in a better HTAP performance. Note that MVCC-based systems~\cite{empiricalMVCC2017} can implement various isolation levels including \textit{read committed}, \textit{repeatable read}, \textit{snapshot isolation}, and \textit{serializable snapshot isolation}. The concurrency control techniques should be different when it comes to different isolation levels. As a result, HTAP techniques may vary slightly because the delta table should adapt to different isolation levels. As similar techniques of concurrency control in OLTP can be applied to HTAP systems, existing HTAP techniques did not discuss these issues.


\textit{Pros and Cons.} MVCC-based HTAP processing has a high freshness as the analytical queries can access the latest visible data by traversing the version chains. However, when it comes to the long version chains, it will incur a significant overhead for version traversing and cleaning.

\subsubsection{Copy-on-Write based HTAP}
This category of techniques~\cite{H2TAP2017, Hyper2011, P-tree2019} relies on in-memory techniques with the Copy-on-Write (CoW) mechanism to support HTAP. The basic idea is to create snapshots for handing OLAP workloads when the main storage encounters a write operation. As shown in Figure~\ref{fig:HP-CoW}(a), a snapshot is generated by forking the main OLTP process with a consistent virtual memory.

For transaction processing, the main OLTP process uses a single thread to execute the transactions in a sequential way without locking and latching~\cite{Hyper2011}, resulting in lock-free transaction processing. Since the data is maintained in the main memory, the OLTP process can execute transactions at the rate of tens of thousands per second. Moreover, CoW can also handle transactions using multiple cores~\cite{P-tree2019} or multiple threads~\cite{Hyper2011} where each thread processes the sequential transactions over a data partition.

For analytical processing, the forked OLAP process handles the queries with the up-to-date snapshots. Particularly, the OLAP process can utilize the OS-enabled interface to create the updated pages on demand, which only takes several microseconds. With multiple threads, the queries can be performed in parallel against a single snapshot or multiple snapshots~\cite{Hyper2011}. Besides, the OLAP process can utilize GPUs to accelerate the query processing~\cite{H2TAP2017}.

\textit{Pros and Cons.} This type of technique enjoys high freshness as the OLAP process can always analyze the fresh data via snapshotting. However, it suffers from large memory size because each process will create a new snapshot, especially for the write-heavy workload. What is more, it has low isolation due to the shared resources in the same instance. 

\subsubsection{Dual-Store based HTAP}
This type of techniques~\cite{Janus2017, saphana_2012,TiDB2020, RateupDB2021,Oracle2015, BatchDB2017, MySQLHeatWave2021,  PostgreSQL2015} relies on a dual-store architecture to handle HTAP workloads. Most of HTAP databases introduced in Section~\ref{sec:HTAP-architecture} employ such an approach by developing a dual-store with both row and columnar format~\cite{ TiDB2020,Oracle2015,MySQLHeatWave2021, PostgreSQL2015}, or purely row format~\cite{BatchDB2017} and purely column format~\cite{saphana_2012,RateupDB2021, Hyper_column}. As shown in Figure~\ref{fig:HP-CoW}(b), a main store is used for handling OLTP workloads, and a secondary store is employed for processing OLAP workloads. The recently updated data is synchronized from the main store to the secondary store in batches periodically and asynchronously.

For transaction processing, the main store relies on MVCC protocols~\cite{empiricalMVCC2017} to process the transactions. Specifically, each insert is first written to the log and the row store, then is appended to the delta store. An update creates a new version of a row with a new lifetime of a begin timestamp and an end timestamp, and the older version is marked as a delete row in a delete bitmap. 

For analytical processing, queries are performed using column-oriented techniques~\cite{ColumnDBMSs} such as compression-aware processing~\cite{DataBlocks}, single-instruction multiple-data (SIMD) instructions~\cite{Oracle2015}, and vector processing~\cite{Hyper_column}. Queries can also be accelerated with new hardware such as heterogeneous CPU/GPU processors, Processing-in-memory (PIM) chips~\cite{Polynesia_ICDE2022, 
lufluidkv}, and FPGAs~\cite{RelationalMemory_EDBT2023}. To analyze the fresh data, the query engine traverses the delta data~\cite{Oracle2015, SQLServer2015}, or uses a merge-on-read mechanism~\cite{TiDB2020,F12020} that merges the delta data to the main store before analyzing the whole data.

\textit{Pros and Cons.} Dual-store HTAP processing has high isolation as the workload can be processed with two separated stores using either logical isolation~\cite{Oracle2015, SQLServer2015, BatchDB2017, RateupDB2021, AlloyDB2022} or physical isolation~\cite{TiDB2020, MySQLHeatWave2021}. However, it has a high synchronization cost as the delta data shall be converted and merged to the secondary store.


\subsection{Data Organization}

Data organization in HTAP databases requires choosing an optimized data layout, e.g., row-wise or column-wise data layout for the mixed workload. As maintaining two copies of data for OLTP and OLAP workloads is costly, existing methods seek to strike a trade-off between the throughput and storage cost. They fall into two categories of approaches: primary row store with selected column store and adaptive hybrid data storage.

\subsubsection{Primary Row Store with Selected Column Store } \label{subsec:column_section}
The first category of methods persists the transaction data in the row store and chooses part of the attributes to be included in the column store. The major objective is to select the beneficial columns (i.e., having high utility of query acceleration and low update cost) into the memory from the primary row store. With such an approach, OLTP workloads are handled by the row store, and OLAP workload can be accelerated by in-memory columnar scan. It has two main methods: (1) frequency-based heatmap and (2) cost-based linear programming. 

\textbf{(1) Frequency-Based Heatmap}. The heatmap approach~\cite{Oracle21c} selects the columns based on the access frequency from the workload. Particularly, the frequently-accessed columns will be kept in the memory and rarely-accessed columns will be evicted to disk. Basically, it groups the in-memory columns into three clusters based on their access frequencies: hot, candidate, and cold columns. The candidate columns are the columns of interests. If some of them were frequently-accessed, then they would be marked as hot columns, which will be populated from the persistent row store. The columns are marked as cold if they were not touched during a time window, e.g., several days. If the cold columns have been populated, they are compressed and evicted to the persistent columns. As extracting the columns from the row store is expensive, the evicted columns can be loaded back if accessed later. 

\textit{Pros and Cons.} The pros of heatmap approach is that it supports arbitrary queries and is easy to implement. The downside is that it may have a low utility as it only considers the access frequency and neglects the effect of various column combinations concerning different queries.

\textbf{(2) Cost-Based Linear Programming}. Another approach is the cost-based linear programming~\cite{hyrise_columnselection}, which formalizes the in-memory column selection as a knapsack-style problem, and then uses an integer linear programming (ILP) method to solve the problem. Given a set of queries, its objective function is a cost function that sums over the scan cost of each query over the involved columns. The goal is to minimize the cost function with a set of columns subject to a given constraint, e.g., the total column size is not greater than a memory budget.

\textit{Pros and Cons.} The ILP method can have high utility because the selected columns can reduce the cost of the queries. However, it is unclear how the cost functions can estimate the cost of complex queries that involve multiple tables and complex operations.



\subsubsection{Adaptive Hybrid Data Storage}
This line of works~\cite{sigmod/proteus,H2O2014,Peloton2016,Casper2019} organizes the data adaptively based on the given workloads and designed cost functions. Such approaches adopt a fine-grained hybrid storage scheme. For example, H2O~\cite{H2O2014} supports three storage layouts: columnar layout, row-wise layout, and column groups (i.e., width-varying vertical partitions of the tables). Given a query workload, it evaluates the cost of different storage schemes (including processing cost and conversion cost), and then selects an optimal storage solution. Casper~\cite{Casper2019} selects an optimal layout of columns based on the mixed workload of read-only queries and updates. It considers various column features, including the number of partitions, the size and range of each partition, the sorting method (i.e., sorted or unsorted), the updating method (e.g., in-place or out-of-place), and the cache size. By evaluating the cost of each scheme based on the defined cost function, it obtains a solution that minimizes the cost with the SLA-aware constraints. Peloton~\cite{Peloton2016} uses a flexible schema to logically divide a relational table into different tiles, where each tile is a sliced data block with vertical and horizontal partitions. With a clustering method, it physically materializes the tiles on disk based on the workload. Beyond the cost-based data organization in a standalone system, Proteus~\cite{sigmod/proteus} leverages machine learning methods to select the optimal mixed row/column storage based on a distributed architecture. It considers a larger design space for the storage, including data format (row-wise or column-wise), data placement (partial or full copy on a node), data compression, and data tiering (memory or disk). Then, it characterizes the workload concerning the storage layout. Finally, it selects a generated storage schema through a learned cost model. Its follow-up work, Tiresias~\cite{Tiresias}, further extends Proteus to support automatic indexing.

\textit{Pros and Cons.} The adaptive hybrid data layout has a lower storage cost and a higher throughput. However, such methods have two main drawbacks. First, the hybrid storage increases the system's complexity in query processing as many execution rules need to be re-implemented. Second, the transaction processing over hybrid storage leads to frequent random accesses, resulting in multiple disk I/Os.

\subsection{Data Synchronization} \label{sec:DS}
As data resides in multiple replicas, efficiently synchronizing the latest transaction data, i.e., the deltas, to the read replicas is required. In addition, since some read replicas adopt a columnar format, it is important to have a tailored method to merge the deltas to the column store efficiently. To address such a problem, two kinds of synchronization methods are proposed. Namely, in-memory delta merging and log-based delta merging.

\subsubsection{In-Memory Delta Merging} \label{subsec:inmemory-delta}
This type of technique synchronizes the data between the row store and column store using the in-memory delta merging. Particularly, there are three key techniques: (1) threshold-based merging; (2) delete-table merging; and (3) dictionary-based merging.

\textbf{(1) Threshold-Based Merging.} This technique periodically merges the delta data to the column store, which has two steps. First, data updates (inserts/updates/deletes) are recorded in a delta table that is normally implemented using a heap table. Second, the delta data will be migrated to the column store when its size reaches a threshold. Particularly, it can use a trickle-based mechanism that constantly migrates the data in the background. 

\textit{Pros and Cons.} This method supports fast insertion as the updates can be inserted into the heap table quickly. However, when the threshold has not been hit, it could slow down the query processing. This is because the analytical query will not only scan the columnar data (i.e., columnar scan) but also traverse the in-memory delta data that has not been merged in order to access the fresh data. Therefore, it has a larger delta traversing overhead since the heap table is unordered, and the analytical query requires a full-table scan to access the heap table.

\textbf{(2) Delete-Table Merging.} This method depicts the delete table-based merging~\cite{SQLServer2015}, which periodically merges the delta data to the column store based on a delete table; the delta store is an index-organized table (i.e., B-tree) that maintains the latest transaction data. The delete table is a bitmap that holds the row IDs (RIDs), where each one indicates a row's location in the column store. The delta merging consists of two phases. In the first phase, it assigns a RID to each row from the delta store and inserts it into the delete table. Then, it transforms the row-wise delta data to the column store with the RIDs that are hidden by the delete table. In the second phase, it removes the RIDs from the delete table, and truncates the data in the delta store. Note that the first phase and the second phase are committed as a transaction, respectively. This is mainly for ensuring the data consistency.


\textit{Pros and Cons.} Delete-table merging supports fast lookup as the delta data is indexed. However, inserting the data has additional overhead due to the data insertion to the index and deleting the table.

\textbf{(3) Dictionary-Based Merging.} This method shows the dictionary-based merging~\cite{SAPHANA2012}, which organizes the delta data in a columnar format and merges the delta to the primary column store based on dictionaries. Particularly, it organizes the delta column by column, and maintains a dictionary with a data vector for each column. The delta merging also consists of two phases. Firstly, new data is merged to a delta column store with a local dictionary and a data vector. Secondly, the local dictionaries are merged into a primary column store with a global dictionary and a global data vector. 


\textit{Pros and Cons.} This method has high efficiency as both the delta store and primary store are organized in a columnar format, and data is indexed using dictionaries. However, as each delta column is indexed by a local dictionary, the volume of delta data grows drastically when the number of data updates increases.




\subsubsection{Log-based Delta Merging}
This type of technique records the delta data in the change log, and then synchronizes the data between the row store and column store with log shipping and replaying. There are two kinds of log-based merging techniques: (1) multi-level delta merging; and (2) change data capture mechanism;

\textbf{(1) Multi-Level Delta Merging.} This method~\cite{TiDB2020, F12020} merges the deltas in the memory level and the deltas in the disk level. It contains four levels from a top-down view: B+tree, memory level, delta space, and persistent storage. Firstly, data manipulation language (DML) operations, such as insert, delete, and update operations, are inserted into a B+ tree after committing the write-ahead log. Secondly, the write operations of a batch are appended to a small delta in the memory, thus the merging process can be performed in batches. Thirdly, the small deltas are compacted and merged into larger delta files in the disk. These small deltas are merged together in the order in which they were written, so these delta files are unordered. Nevertheless, the multi-version, duplicate, and rolled-back records will be removed in the process of merging. Lastly, the unordered delta files will be periodically merged into the persistent storage in a columnar format. Particularly, the persistent storage organizes the data with the ordered chunks, where each one covers a part of the range of the data. Since the delta files are out of order, the merge operation will produce a large overhead. Therefore, it will locate the data by searching over the B+ tree, and then merge the data with the order.

\textit{Pros and Cons.} Multi-level delta merging has high scalability as the delta data is organized in multiple stages with partitioned files. However, it has a high merging cost due to the large I/O overhead.  

\textbf{(2) Change Data Capture (CDC).}  CDC is another merging mechanism~\cite{IDAA_VLDB2020,  SAPHANA2017,MySQLHeatWave2021}, which monitors the data updates in the change log, and replicates the change log to the analytical store asynchronously. Generally, it treats the log as the first citizen, and then migrates the valid log records to the column store.


\textit{Pros and Cons.} The CDC mechanism has a high-performance isolation as the OLTP and OLAP workloads are physically isolated in different nodes or systems. However, it incurs high latency due to the log shipping and replaying.


\subsection{Query Optimization}
We summarize three types of query optimization techniques: (1) hybrid row/column scan \cite{TiDB2020, SQLServer2015}; (2) multi-version indexing~\cite{P-tree2019, MVPBT2020} and (3) CPU/GPU acceleration \cite{RateupDB2021, H2TAP2017}. Next, we dive into each type of technique.

\subsubsection{Hybrid Row/Column Scan}
In the HTAP databases, a complex query can be routed to perform against either the row store or the column store. We call such an execution mode as hybrid row/column scan~\cite{TiDB2020, SQLServer2015, Oracle21c}. Moreover, a query can also be executed in a fine-grained way such that part of the operators are processed in the column store, and the rest of the operators are processed in the row store, and finally the results are combined. For instance, the short-range or point queries can be performed using B+ tree indexes in a row store; column scans and complex aggregations can be processed using SIMD scans in a column store; the query coordinator merges the results from both execution engines to the final results. Suppose a SQL query finds the license and color of the vehicles registered in Beijing as follows:

\vspace{1mm}
SELECT V.license, V.color 

FROM Register R, Vehicle V 

WHERE R.VID=V.ID and R.place ="BJ"
\vspace{1mm}

Such an SQL query contains a two-way join between the Register and Vehicle tables with an equality predicate on the place field. The logical plan is separated into a hybrid plan, which relies on a B+ tree to search for the qualified records in the Register table, then joins their VIDs with the IDs of the Vehicle table in the column store, finally it returns the results by projecting the columns of license and color in the Vehicle table. Hence, the query execution can benefit from both the index scan in the row store and the columnar scan.


Various interfaces for hybrid scan~\cite{AlloyDB2022, Oracle2015, SQLServer2015, TiDB2020} have been developed. For instance, Oracle~\cite{Oracle2015} can create an in-memory columnar table by altering the table with "INMEMORY" keyword, then any given SQL query can be executed with a rule-based hybrid scan. SQL Server~\cite{SQLServer2015} supports the hybrid scan by building a "COLUMNSTORE INDEX" over the target attributes or tables, and then it accelerates the queries with cost-based columnar scans. AlloyDB~\cite{AlloyDB2022} performs a cost-based hybrid scan in the operator granularity after enabling its columnar engine. TiDB~\cite{TiDB2020} allows for creating $n$ columnar replicas by altering a table T with "SET TIFLASH REPLICA n", then the queries can be executed using distributed columnar scans. It also supports the hint-based hybrid scan, which can force the access paths using hints. Recall the above SQL query. If a hint "read\_from\_storage(TIKV[Register], TIFLASH[Vehicle])" is added to the SQL query, then the Register table will be probed from the TIKV row store and the Vehicle table will be scanned from the TIFLASH column store.

The key to hybrid row/column scans is to determine whether a query or an operator should be executed against the row or column store. However, it is not always straightforward to generate an optimal plan for a more complex query to which the plan space is large.  Existing methods are mainly grouped into two types: (1) rule-based execution; and (2) cost-based execution.


\textbf{(1) Rule-based Execution.} This type of method utilizes heuristic rules to execute the queries. They rely on two rules of thumb, namely, (i) the columnar scan is more efficient than the index scan and row scan; and (ii) the index scan is more efficient than the row scan. For instance, Oracle~\cite{Oracle21c} performs the hybrid scan by following a "column first, row later" principle: if some columns do not exist in the column store, it scans them in the row store, and finally merges the results from the columnar scan and row scan. 

\textit{Pros and Cons.} The rule-based methods have a high efficiency because of the short planning time by the rules of thumb. However, they may miss the optimal plan as they do not explore the global plan space. For example, an index row scan may be more efficient than a columnar scan, depending on the cost of the specific plans. Therefore, it is preferable to consider the cost of candidate plans as well.




\textbf{(2) Cost-based Execution.} This type of methods~\cite{TiDB2020, SQLServer2015, Polardb2021} selects the access path by comparing the cost of the candidate execution paths. For instance, TiDB~\cite{TiDB2020} builds a cost model among the columnar scan, row scan, and index scan, and it selects the access path with the minimum cost. Unfortunately, the cost model considers only scan operations. PolarDB-IMCI~\cite{polar-IMCI} utilizes a threshold-based cost model to select either the row scan or columnar scan. It relies on the row-based cost model, and if a query cost is beyond a pre-defined threshold, the query is routed to the column store. Metis~\cite{mentis} is a recent work that can generate HTAP-aware hybrid plans by considering the cost of data updates and data synchronization. Its core contribution is a new cost model that considers the delta scan, columnar scan, index scan, and row scan, which can guide the access path selection.


\textit{Pros and Cons.} This method can produce query plans of high quality. However, existing cost functions are based on independence and uniform distribution assumptions, so the estimation may be inaccurate when these assumptions do not hold. Besides, existing works lack a global and comprehensive cost model for HTAP workloads.









\subsubsection{Multi-Version Indexing}
Multi-version indexing aims to accelerate HTAP through new indexing methods. Two representatives are Parallel Binary Tree (P-Tree)~\cite{P-tree2019} and Multi-Version Partitioned B-Tree (MV-PBT)~\cite{MVPBT2020}. 



The main idea of P-Tree is to replicate the data paths involved in the latest transaction on the balanced binary tree, and it leverages multi-core processors to operate on the data concurrently. Furthermore, P-Tree implements the snapshot isolation level, so queries can also access snapshot data visible on the index at the same time. The read operation obtains the pointer of the root node, and then follows the index to find the visible versions of data on the path; the update operation will create a new version of the root node, then copies all relevant paths and updates the target node; both read and update operations can be completed in O(log n) time. P-Tree also supports nested mode across multiple tables, so queries can access cross-table data through indexes without joins. 

MV-PBT utilizes a multi-version partitioned B-Tree to index the updated data versions. The motivation is that there could be long version chains for an MVCC-based HTAP database, so efficient indexing for the version chains is required. The main challenge is how to efficiently index multi-version data of the same data and support fast queries of the latest visible version data. MV-PBT is based on the partitioned B-Tree that is divided based on the specified key, and each partition has the same search key. All updates of the transaction will be written to the memory buffer; when the buffer is full, the data will be persisted to the corresponding partitioned B-Tree. 

\textit{Pros and Cons.} P-Tree has a high efficiency as the data can be read and updated concurrently. However, the downside is that it consumes large memory size and CPU resources as a single update leads to a copied path. Since different versions of the same data have been indexed into the partition B-Tree, it supports a fast search of visible versions of the data. Compared to the P-Tree, MV-PBT has a higher scalability as the data is indexed using a disk-based B-tree. However, it has a low throughput due to the row-based query processing.


\subsubsection{CPU/GPU Acceleration}
The heterogeneous integrated processor of CPU/GPU is also an important technology for HTAP query optimization~\cite{H2TAP2017, RateupDB2021, CPU-GPU-Query}. This type of technique utilizes the task-parallel nature of CPUs and the data-parallel nature of GPUs for handling OLTP and OLAP, respectively. Existing CPU/GPU methods for HTAP are based on a dual store to isolate the workload execution of OLTP and OLAP workloads. That is, using CPU and transactional store for OLTP and using GPU and analytical store for OLAP. Therefore, the kind of technique is also called heterogeneous HTAP ($\mathtt{H^2TAP}$)~\cite{H2TAP2017}. Particularly, the HTAP workload is classified into OLAP workload and OLTP workload; OLAP is executed on the analytical store through the GPU, and OLTP is processed on the transactional store with CPU cores; the data updated by the transactional storage can be synchronized to the analytical storage in batches. Ideally, the analytical store will organize the data in a columnar format, and the transaction store will use a row-based format. However, existing methods only adopt purely column-based format~\cite{H2TAP2017, RateupDB2021} due to the consideration of the engineering complexity. Therefore, this approach favors high OLAP throughput but has a low OLTP throughput. It is worthwhile to mention that the architecture and processing paradigm of $\mathtt{H^2TAP}$ has many variants and can be changed for different types of workloads accordingly. For instance, some CPU cores can also be deployed in the GPU chips to accelerate the short queries in parallel~\cite{H2TAP2017}. Besides, recent work shows that transactions can be accelerated using GPU~\cite{GPU-transactions}, and read-only queries can be executed in both CPU and GPU with a proper data placement strategy~\cite{CPU-GPU-Query}.

\textit{Pros and Cons.} The $\mathtt{H^2TAP}$ method favors high analytical throughput because queries can be accelerated by GPU or hybrid scan of CPU/GPU. However, it suffers from the low freshness issue due to the low network bandwidth of PCIe between CPU and GPU.

\subsection{Resource Scheduling}

HTAP databases need to support the efficient execution of OLTP and OLAP workloads simultaneously, but the performance degradation could be remarkable due to the data synchronization and resource contention. Besides, data freshness is another concern if OLAP cannot read the latest updated data. Existing methods include freshness-driven scheduling and workload-driven scheduling.




\subsubsection{Freshness-Driven Scheduling}
This method~\cite{ResourceScheduling2020} switches the execution modes based on a freshness threshold. Each execution mode adopts a particular strategy for resource allocation and data exchange. For instance, the scheduler controls the execution of OLTP and OLAP in isolation for high throughput, and then periodically synchronizes the data. Once the data freshness becomes low, it switches to an execution mode where OLTP and OLAP share the same copy of data such that the queries can access the fresh data directly.



Such a method works by varying three execution modes: (S1) Co-located execution with the OLTP instance and OLTP instance, where the OLTP instance handles transactions and can access the fresh data with the copy-on-write mechanism; (S2) Isolated execution with the OLTP instance and OLAP instance, where OLTP instance handles transactions and OLAP instance processes the queries, and the delta data in OLTP instance is periodically synchronized to the OLAP instance; (3) hybrid execution with two OLTP instances and the OLAP instance, where OLAP queries need to analyze the base data in the OLAP instance and the delta data in OLTP instance simultaneously. For resource scheduling, the system executes S2 mode by default, which favors high-performance isolation. When the data freshness is less than the given threshold specified in the service-level agreement (SLA), it can jump to S1 or S3. There is a trade-off among S1, S2, and S3: S1 can analyze the fresh data immediately, but the analytical capacity is limited; S3 can analyze the fresh data with more CPU and storage resources, but it needs to access two instances for query processing. S2 handles the hybrid workloads with better isolation, but the data freshness is lower due to the latency of data exchange. 

\textit{Pros and Cons.} The freshness-driven scheduling strikes a good balance between performance and freshness. However, the system performance may fluctuate due to the frequent mode switching. Therefore, it might be helpful to design some mechanisms for lazy switching.


\subsubsection{Workload-Driven Scheduling}
This kind of methods~\cite{TaskScheduling, TPCTC2014, performance_isolation_2021} dynamically schedules resources such as CPU, shared cache, and memory bandwidth, by monitoring the execution of the mixed workload. Since the access patterns of OLTP and OLAP workloads are different, the resource scheduling needs to adapt to their performance characteristics.

For CPU resources, workload-driven scheduling adjusts the parallelism threads of OLTP and OLAP tasks. The initial number of threads of OLTP is set to the number of CPU cores, and the number of threads of OLAP is set to an average parallelism based on the history statistic. The scheduling method adaptively adjusts the number of threads based on the performance of executed workloads. For example, when CPU resource is saturated by OLAP threads, the task scheduler can decrease the parallelism of OLAP while enlarging the OLTP threads. When the monitoring process (e.g., a watchdog~\cite{TaskScheduling}) detects a blocked transaction thread, it will join it to a blocking queue and will try to restart the blocked task. For shared cache and memory bandwidth resources, it can also be dynamically adjusted by observing the workload execution~\cite{performance_isolation_2021}. For example, when the OLAP throughput decreases drastically in a hybrid execution, it indicates that the OLTP execution affects the OLAP execution. Hence, more shared cache, e.g., LLC cache, can be assigned to the OLAP instance. In addition, when the OLAP throughput drops due to the synchronization process, it is necessary to allocate more resources to the synchronization process.

\textit{Pros and Cons.} This method has high throughput as it regards the throughput metric as the first-class citizen. However, it has a low freshness as it does not consider the freshness metric at all.
\section{HTAP Benchmarks} \label{Sec:HTAP-Benchmark}


\begin{table*}[]
\renewcommand\arraystretch{1.15}
\caption{An Overview of HTAP Benchmarks\label{table:benchmarks}}
\begin{tabular}{|c|c|c|c|c|c|}
\hline
Benchmark                  & Type                                                                              & Domain/Task            & Benchmark/Workload                                                & Execution Rule                                                                             & Metrics                                                                               \\ \hline
CH-Benchmark~\cite{CH-benchmark}           & \begin{tabular}[c]{@{}c@{}}End-to-End \\ Benchmark\end{tabular}                   & Retail Business   & \begin{tabular}[c]{@{}c@{}}TPC-C+ TPC-H\end{tabular} & \begin{tabular}[c]{@{}c@{}}Isolated Execution+ \\ Hybrid Execution\end{tabular} & \begin{tabular}[c]{@{}c@{}}Referenced\\ Throughput\end{tabular}                 \\ \hline
HTAPBench~\cite{htapbench}                  & \begin{tabular}[c]{@{}c@{}}End-to-End \\ Benchmark\end{tabular}                   & Retail Business   & \begin{tabular}[c]{@{}c@{}}TPC-C+ TPC-H\end{tabular} & \begin{tabular}[c]{@{}c@{}}Fixed OLTP workers+\\ varied OLAP workers\end{tabular}   & \begin{tabular}[c]{@{}c@{}}Referenced\\ Throughput\end{tabular}                 \\ \hline
\multirow{3}{*}{OLxPBench~\cite{conf/icde/OLxPBench}} & \multirow{3}{*}{\begin{tabular}[c]{@{}c@{}}End-to-End \\ Benchmarks\end{tabular}} & Retail Business   & TPC-C+9 queries+5 txns                                                  & Hybrid Execution                                                                           & Throughput                                                                            \\ \cline{3-6} 
                           &                                                                                   & Banking   & Small bank +4 queries+6 txns                                             & Hybrid Execution                                                                           & Throughput                                                                            \\ \cline{3-6} 
                           &                                                                                   & Telecom           & TATP+5 queries + 6 txns                                                   & Hybrid Execution                                                                           & Throughput                                                                            \\ \hline
HATtrick~\cite{sigmod22/HAPtrick}                    & \begin{tabular}[c]{@{}c@{}}End-to-End\\ Benchmark\end{tabular}                    & Retail Business   & \begin{tabular}[c]{@{}c@{}} SSB + 3 txns\end{tabular}   & Hybrid Execution                                                                           & \begin{tabular}[c]{@{}c@{}}2D Throughput\\ Freshness\end{tabular} \\ \hline

HyBench~\cite{pvldb/HyBench}      & \begin{tabular}[c]{@{}c@{}} End-to-End\\ Benchmark\end{tabular}                    & Online Finance & \begin{tabular}[c]{@{}c@{}} 18 txns+13 queries\\ +12 mixed operations \end{tabular}   & Hybrid Execution                                                                          & \begin{tabular}[c]{@{}c@{}} F-Score\\ H-Score\end{tabular} \\ \hline

ADAPT~\cite{Peloton2016}                      & Micro-benchmark                                                                    & Data Organization & 1 insert + 4 select queries                                         & Hybrid Execution                                                                           & Throughput                                                                            \\ \hline
HAP~\cite{Casper2019}                        & Micro-benchmark                                                                    & Data Organization & 6 CRUD queries                                         & Hybrid Execution                                                                           & Throughput                                                                            \\ \hline
mOLxPBench~\cite{kang2023benchmarking}                        & Micro-benchmark                                                                 & Data Synchronization & 6 CRUD queries                                      & Hybrid Execution                                                                       & Tail Latency                                                                          \\ \hline
\end{tabular}
\end{table*}
Table \ref{table:benchmarks} summarizes eight state-of-the-art HTAP benchmarks, including five end-to-end benchmarks and three micro-benchmarks. 



\subsection{CH-Benchmark}
CH-benchmark~\cite{CH-benchmark}, a.k.a., TPC-CH~\cite{TPC-CH}, is a widely-used end-to-end HTAP benchmark that combines two classical TPC benchmarks, i.e., TPC-C \cite{TPCC} for benchmarking transactional processing systems, and TPC-H \cite{TPCH} for benchmarking analytical reporting systems.

(1) \textbf{Data Schema}. CH-benchmark unifies the schema of TPC-C and TPC-H in an application domain of retail business, simulating the behaviors of wholesale suppliers that process the customers' orders and analyze the fresh sales data simultaneously. Specifically, it combines TPC-C's nine tables and TPC-H's eight tables to a schema of 12 tables by merging the overlapping tables from TPC-H (i.e., customer, orders, lineitem, part) and by removing its partsupp table. It has adjusted the scaling model for data generation based on the number of warehouses.

(2) \textbf{Workloads} Since the schema of TPC-C has not changed, the CH-benchmark preserves all the five transactions of TPC-C for OLTP. For OLAP workloads, it has made several modifications on TPC-H's. First, it has adjusted the tables' names and join keys of the original 22 queries. Second, it has reduced the arithmetic operations in the queries. Third, it has removed the refresh function of TPC-H as TPC-C has included the operations of data updates.

(3) \textbf{Execution Rule}. The execution rule involves two steps: (i) it first executes \textit{n} streams of OLTP workloads and \textit{m} streams of OLAP workloads in isolation, then (ii) it performs the mixed workloads with the same number of streams in parallel. The main purpose is to evaluate how HTAP systems can handle the interference of two types of workloads by comparing the performance between the isolation mode and the hybrid mode.
    
(4) \textbf{Performance Metrics}. CH-benchmark has a reference-based metric to measure performance by combining the metrics of tpmC and QphH. Particularly, tpmC is the number of transactions processed by the system per minute, and QphH is the number of queries handled by the system per hour. Two reference-based metrics are as follows:

\begin{equation}
\label{equ:reference_metric}
\small
    \mathit{M(OLTP)} = \frac{tpmC}{QphH}@tpmC
\end{equation}

\begin{equation}
\small
\label{equ:reference_metric}
    \mathit{M(OLAP)} = \frac{tpmC}{QphH}@QphH
\end{equation}

\noindent where both metrics consist of two parts: the former part measures a ratio between tpmC and QphH, and the latter part is the referenced primary metric, i.e., tpmC for M(OLTP) and QphH for M(OLAP). Suppose M(OLTP) is used, and the execution rule follows the isolation-and-hybrid execution mode. For the isolated/sequential execution, M(OLTP) is 2.5@5000 tpmC. Meanwhile, for the hybrid execution, M(OLTP) becomes 3@5500 tpmC, indicating the OLTP throughput is increased with hybrid execution. 

\subsection{HTAPBench}
HTAPBench \cite{htapbench} also combines TPC-C and TPC-H. It adopts the same data schema and the same hybrid workloads in CH-benchmark. Its main contribution is two-fold. First, it curates the parameters of OLAP queries with the concept of a dynamic window. Second, it contains an execution rule for targeting a fixed OLTP throughput by adaptively launching the OLAP workers.   


(1) \textbf{Workloads}. HTAPBench employs the mixed workload of TPC-C and TPC-H. However, since TPC-H adopts the fixed parameters for the queries, the performance results across runs are often incomparable in the hybrid execution mode. This is mainly because TPC-C's workload may update different parts of data during execution and change the data distribution. To address such a problem, HTAPBench leverages the dynamic query generator, which utilizes a density consultant to generate the parameters for the analytical queries to ensure the same query selectivity during execution. Particularly, it curates the parameters based on the DATE field to slide the time windows of queries for accessing the newly inserted data.

(2) \textbf{Execution Rules}. Instead of scaling the query streams of OLTP and OLAP simultaneously, HTAPBench regards the OLTP throughput as the first citizen. That is, it targets a fixed OLTP throughput and evaluates how well the SUT systems process the OLAP workloads while keeping the OLTP throughput and scaling the OLAP workers. Specifically, according to the TPC-C's specification that defines the maximum of 1.286 tpmC per client, it computes the number of clients and warehouses for a target tpmC, then populates the databases and launches the OLTP workload execution. Then, it gradually starts the OLAP workers based on a client balancer, which monitors the throughput gap between the actual tpmC and the target tpmC, and periodically decides whether or not to start an additional OLAP worker. Namely, if the current gap is larger than a margin and the resource is not saturated, then a new OLAP worker will be started. Otherwise, the balancer keeps monitoring.   
    
(3) \textbf{Performance Metrics}. Based on the execution rule, HTAPBench proposed a OLTP-oriented metric for measuring the HTAP performance as follows:

\begin{equation}
\small
\label{equ:target_metric}
    QpHpW = \frac{QphH}{\#OLAPworkers}@tpmC
\end{equation}

\noindent where QpHpW denotes the number of processed Queries per Hour per Worker. The former part is a ratio between TPC-H’s metric and the total number of OLAP workers. The latter part is the target tpmC. 


\subsection{OLxPBench}
OLxPBench~\cite{conf/icde/OLxPBench} is a composite HTAP benchmark suite. It consists of three domain-specific HTAP benchmarks inherited from three established OLTP benchmarks, respectively. Namely, \textit{Subenchmark} is from the TPC-C benchmark~\cite{TPCC} at a retail business scenario, \textit{Fibenchmark} is from the SmallBank~\cite{SmallBank} benchmark at a bank scenario, and \textit{Tabenchmark} is from the TATP~\cite{TATP} benchmark at a telecom scenario. OLxPBench makes two modifications. First, it contains new analytical queries with the transactions to compose an HTAP workload. Second, it includes analytical transactions that perform real-time queries inside the transactions.

(1) \textbf{Data Schema}. The OLxPBench benchmarks have the schema of three original OLTP benchmarks. Particularly, \textit{Subenchmark} preserves nine tables from TPC-C; \textit{Fibenchmark} keeps three tables from SmallBank, i.e., ACCOUNT, SAVING, CHECKING; \textit{Tabenchmark} includes all the five tables from TATP, which simulates a Home Location Register (HLR) database used by a mobile carrier.
    
(2) \textbf{Workloads and Execution Rules}. Unlike CH-benchmark where the analytical queries only operate on the overlapping tables of TPC-C, OLxPBench follows semantic consistency, meaning that both queries and transactions shall cover all the tables. For instance, the history, warehouse, and district tables will never be touched in the queries of CH-benchmark, while OLxPBench contains several queries that analyze the records from these tables. Moreover, OLxPBench includes analytical transactions that are originally proposed by Gartner, which envisioned HTAP transactions that include analytical operations. Specifically, OLxPBench implements such a transaction by adding a real-time query to each transaction in all the three benchmarks. With the new workload, OLxPBench offers three execution modes: (1) a sequential execution mode of transactions and queries. (2) a concurrent execution mode of mixed workloads. and (3) an execution mode of analytical transactions.

\subsection{HATtrick}
HATtrick~\cite{sigmod22/HAPtrick} is an end-to-end HTAP benchmark, which combines an analytical benchmark, SSB~\cite{tpctc2009/SSB}, with a transactional component inspired by TPC-C. It has two contributions to the performance metrics. First, it proposes the metric of the throughput frontier to capture the OLTP and OLAP throughput with a 2D visualization graph. Second, it uses the freshness score to quantify the transaction's recency for a query by the analytical client (A-client).

(1) \textbf{Data Schema}. HATtrick modifies the SSB schema from two aspects. First, it adds a HISTORY table and updates the relations of CUSTOMER, SUPPLIER, and PART by adding new attributes that will be used by the transactions. Second, it incorporates a set of entries $\mathit{FRESHNESS}_j$, where each entry stores an integer transaction number TXNNUM for the $i$-th transactional client (T-client). TXNNUM is used to calculate the freshness score for the queries.
    
(2) \textbf{Workloads and Execution Rules}. For analytical queries, HATtrick contains all 13 queries of SSB with a modification to return the freshness entries. Each A-client performs a batch of 13 queries recursively with random orders. For transactions, HATtrick defined three transactions based on TPC-C, including two read-write transactions, NEW ORDER and PAYMENT, and a read-only transaction, COUNT ORDERS. Each T-client defined the ratio of transactions with 48\%, 48\%, and 4\%, respectively. Each transaction committed by the T-client $j$ will update the $\mathit{FRESHNESS}_j$ with the transaction ID accordingly. By configuring the number of T-clients and A-clients, HATtrick executes the workloads concurrently.

(3) \textbf{Performance Metrics}. HATtrick introduces two metrics: throughput frontier and freshness score. Throughput frontier is a 2D graph where the x-axis represents the throughput of T-clients, and the y-axis depicts the throughput of A-clients. By fixing either the T-clients or A-clients while varying the number of the other types of clients, the throughput frontier can be effectively computed. The plotted throughput frontiers can reflect the relationships between performance characteristics and workload interference. For instance, if the frontier is below the diagonal line of the bounding box, then the hybrid throughput is relatively low, and workload contention is high. If the frontier is close to the bounding box, then the hybrid throughput is high, and the performance isolation is good. With a global clock, HATtrick defined the freshness score $f_{A_q}$ of an analytical query $A_q$ as follows:

\begin{equation}
\small
\label{equ:freshness_metric}
    f_{A_q} = max(0, t^s_{A_q} - t^{f_{ns}}_{A_q})
\end{equation}

\noindent where $t^{f_{ns}}_{A_q}$ is the commit time of the first transaction $t^{f_{ns}}$ that is unseen by $A_q$. $t^s_{A_q}$ is the start time of $A_q$. Intuitively, the smaller the score is, the fresher the data is. The larger the score is, the analyzed data is more stale.

\subsection{HyBench}
As existing benchmarks~\cite{htapbench, CH-benchmark, conf/icde/OLxPBench, sigmod22/HAPtrick} heavily rely on traditional OLTP benchmarks (e.g., TPC-C) or OLAP benchmarks (e.g., TPC-H), they fall short of providing representative HTAP data, workload, and metrics.

HyBench~\cite{pvldb/HyBench} is a newly-emerged benchmark for HTAP databases,   which features a new data generator, a multi-set workload, and a unified metric. First, it contains a schema by simulating a realistic online finance HTAP scenario, and it provides a data generator based on a time-dependent generation phase and an anomaly generation phase. Regarding the workload, it has three sets of workloads for OLTP, OLAP, and OLXP, evaluating the performance of transaction processing, analytical processing, and hybrid processing, respectively. To quantify the overall HTAP performance, it proposes a unified metric, H-Score, that combines the performance of OLTP (TPS), OLAP (QPS), and OLXP (XPS) and data freshness.

(1) Data Schema. Its schema is based on \marginpar[\rev{}]{} an online finance application inspired by the real-world HTAP applications in the field of finance technology (FinTech)~\cite{cikm/account_detection, WeBank1, OceanBase}. The schema consists of eight tables, including CUSTOMER, COMPANY, SAVINGACCOUNT, CHECKINGACCOUNT, TRANSFER, CHECKING, LOANAPP, and LOAN, simulating the widely-used finance activities such as saving, payment, and loan application. The data generation produces the testing data based on a given scale factor (SF), and the data size grows linearly as the SF increases. Instead of using uniform data generation~\cite{sigmod22/HAPtrick,fan2020relational}, HyBench leverages a time-dependent data generation to generate data in three time ranges, enabling efficient and realistic data generation. Additionally, it proposes an anomaly generation phase to produce blocked accounts and illegal transactions, which simulate realistic anomalies.

(2) Workloads and Execution Rules. HyBench contains 18 operational transactions, 13 analytical queries, and 12 mixed operations that include six analytical transactions (AT) and six interactive queries (IQ), providing a rich set of workloads for benchmarking HTAP databases. For instance, AT1 makes a transfer while performing the risk controlling by analyzing if the target has any risks; IQ1 finds related transfers for a blocked user on-the-fly. HyBench follows a three-phase execution rule to execute the hybrid workload in a row, including an AP phase, a TP phase, and an XP phase, and then it outputs a unified metric.

(3) Performance Metrics. HyBench proposes two new metrics tailored to HTAP databases, namely, F-Score and H-Score. F-Score is used to measure the data freshness by measuring the timestamp difference between the result set from the OLTP instance and the OLAP instance. Particularly, it is a general method that supports the cases of insert, update, and delete. H-Score is a unified metric that relies on geometric mean to measure the overall HTAP performance. Given the concurrency of TP workers and AP workers ($n,m$), it is defined as follows:

\begin{equation}
\label{equ:unified_metric}
    \texttt{H-Score} = SF* \frac{\sqrt[3]{TPS * QPS * XPS}}{\overline{f_s}+1}
\end{equation}

\noindent where XPS = ATS+IQS, and ATS and IQS are analytical transactions per second and interactive queries per second, respectively; SF is the used scale factor; $\overline{f_s}$ is the F-Score that is measured in seconds, and it is added with 1 to avoid that the denominator is zero.

\subsection{Micro-Benchmarks}
Other than the end-to-end benchmarks, there are three synthetic micro-benchmarks~\cite{Peloton2016, Casper2019, kang2023benchmarking} that were developed to evaluate the adaptive data layout in HTAP databases. Particularly, ADAPT~\cite{Peloton2016} contains two synthetic tables with 50 and 500 integer attributes, respectively. Each table owns a primary key $a_0$ and has ten million tuples. The workload operates on either the narrow table or the wide table, i.e., the table with 500 attributes. It contains an insert query, three selection queries, and a self-join query. The selection queries project a subset of the attributes ({$a_1,\dots a_p$}) with a filter on the primary key $a_0$. The self-join query projects a subset of the attributes with a theta-join on two random attributes, i.e., $X.a_i<Y.a_j$. HAP~\cite{Casper2019} was inspired by ADAPT. For the data schema, it contains a narrow table of 16 columns and a wide table of 160 columns. Regarding the workload, it contains six queries, including a point query with a projection, two aggregation queries with range filters, an insert query, a delete query, and an update query with an equality filter on the primary key.  mOLxPBench~\cite{kang2023benchmarking} is a new HTAP microbenchmark, which contains a single ITEM table with 59 attributes and includes five queries and one update. A salient feature is that it can control the rate at which fresh data is generated and the scan range. Therefore, the greater the rate and range are, the more data needs to be synchronized, and the tail latency increases sharply.
\section{Open Problems and Challenges}

\noindent \textbf{Data Organization for Distributed HTAP Databases.} As various HTAP databases are going towards distributed architectures, the data organization poses many new challenges. The first challenge is to decide which columns to keep in memory, which on disk, and on which nodes. The second challenge is to decide whether the data is organized in a row-wise or column-wise or unified format~\cite{NoisePage_Arrow, L-store2016}. The third challenge is to decide which compression methods~\cite{Hyrise_encoding} should be used, and in which granularity (table or column or segment~\cite{Hyrise2019}). Proteus~\cite{sigmod/proteus, Tiresias} is a recent representative that employed an offline learning method considering many factors to organize the data. However, the offline learning method needs to be further justified due to the high complexity of model learning and data organization. It might be helpful to address these challenges separately to reduce the complexity. In addition, combining the offline learning with lightweight online learning methods~\cite{ICDE2022-ML4AL, li2021ai, conf/aiml2/LiZC21, zhang2023autoce, zhang2024pace, sun2021learned} can also mitigate the training overhead.


\noindent \textbf{Query Optimization in HTAP Databases.} There are several open problems for HTAP query optimization. The first one is about hybrid scans for analytical queries. As the existing interface has limited functionality for hybrid scans (e.g., data cannot be exchanged between the row and column data), it calls for more flexible and effective methods to generate hybrid plans. The second open problem is the FPGA-enabled HTAP. Recent works~\cite{RelationalMemory_EDBT2023, NDP_HTAP} have shown some promising results on such a task, but more HTAP workloads need to be explored as many operators have yet to be implemented. Finally, how to incorporate the learned indexes~\cite{LearnedIndex,journals/pvldb/ML4DB21} for HTAP is also an open issue.

\noindent \textbf{Holistic Scheduling for HTAP.} Existing freshness-driven scheduling \cite{ResourceScheduling2020} relies on a rule-based approach to control the execution modes with different freshness settings. The workload-driven approaches \cite{SAPHANA2012, performance_isolation_2021} only adjust the number of OLTP and OLAP threads but do not consider the freshness. Consequently, there still lacks a holistic scheduling method that can orchestrate the workloads, resources, and freshness together. For example, if the current delta store is too cumbersome, some OLAP queries may be scheduled to OLTP instances for high freshness. Therefore, it is preferable to develop a holistic scheduling method that not only captures the workload pattern for better performance, but also satisfies the requirements of data freshness and cost.

 \noindent \textbf{HTAP for Multi-Model Data Analytic.} As other data models such as graph models are calling for the support of HTAP~\cite{graphHTAP}, it is also promising to enable HTAP for multi-model data analytics. Gart~\cite{gart} is a pioneering work that supports the HTAP over row store, and then synchronizes the delta logs to the graph store. Nevertheless, there still remain many opportunities, such as supporting HTAP-aware multi-model queries~\cite{UniBench_journal, UniBench, guo2024multi} and supporting other data models like semi-structured document~\cite{zhang2020selectivity}.

\noindent \textbf{Serving atop HTAP.} Building data services on top of the HTAP databases is an interesting direction to enable freshness-driven data serving including real-time machine learning (ML) based data analytics. Such a concept is also called HSTAP~\cite{kang2023nhtapdb, HSTAP}, and how to efficiently and effectively train the ML over the incremental transaction data remains unexplored.

\noindent \textbf{Cloud-Native HTAP Techniques.} Cloud-native HTAP techniques~\cite{AlloyDB2022, SingleStore2021, Snowflake2022} are just unfolding and bring many new challenges for HTAP. First, since the compute and storage are disaggregated, it is challenging to deliver a high data freshness if the log in the storage layer has not been replayed. Thus, how to guarantee the data freshness with a low replication latency for the compute layer is an open problem. Second, it is challenging to schedule the OLAP and OLTP workloads to meet various requirements of multiple tenants (e.g., throughput or freshness or cost). Hence, it calls for SLA-aware HTAP scheduling methods. Third, as serverless computing~\cite{journals/ftdb/NarasayyaC21, conf/sigmod/Starling20, journals/cacm/ServerlessComputing21} is becoming prevalent, it is still an open problem to utilize serverless computing to handle the HTAP workloads.



\section{Conclusion}
In this paper, we review the recent advancement of HTAP databases. We classify the state-of-the-art HTAP databases according to four storage architectures. We compare their pros and cons, summarize the challenges and opportunities, and discuss the suitable applications. Since ``one HTAP database cannot fit all", we recommend choosing different HTAP architectures to meet the requirements of specific applications. Furthermore, we present their key techniques regarding hybrid workload processing, data organization, data synchronization, query optimization, and resource scheduling, we then summarize the pros and cons of various techniques. We also compare and summarize the state-of-the-art HTAP benchmarks, concerning domain applications, data schema, workload, execution rules, and metrics. Finally, we discuss the research challenges and open problems for HTAP techniques.


\section*{Acknowledgments} 
This paper was supported by the National Key R\&D Program of China (2023YFB4503600), NSF of China (61925205, 62232009, 62102215), Huawei, BNRist, CCF-Huawei Populus Grove Challenge Fund (CCF-HuaweiDBC202309). Guoliang Li is the corresponding author.


\bibliographystyle{abbrv}
\bibliography{main}

\vspace{-3 em}

\begin{IEEEbiography}[{\includegraphics[width=1in,height=1.25in,clip,keepaspectratio]{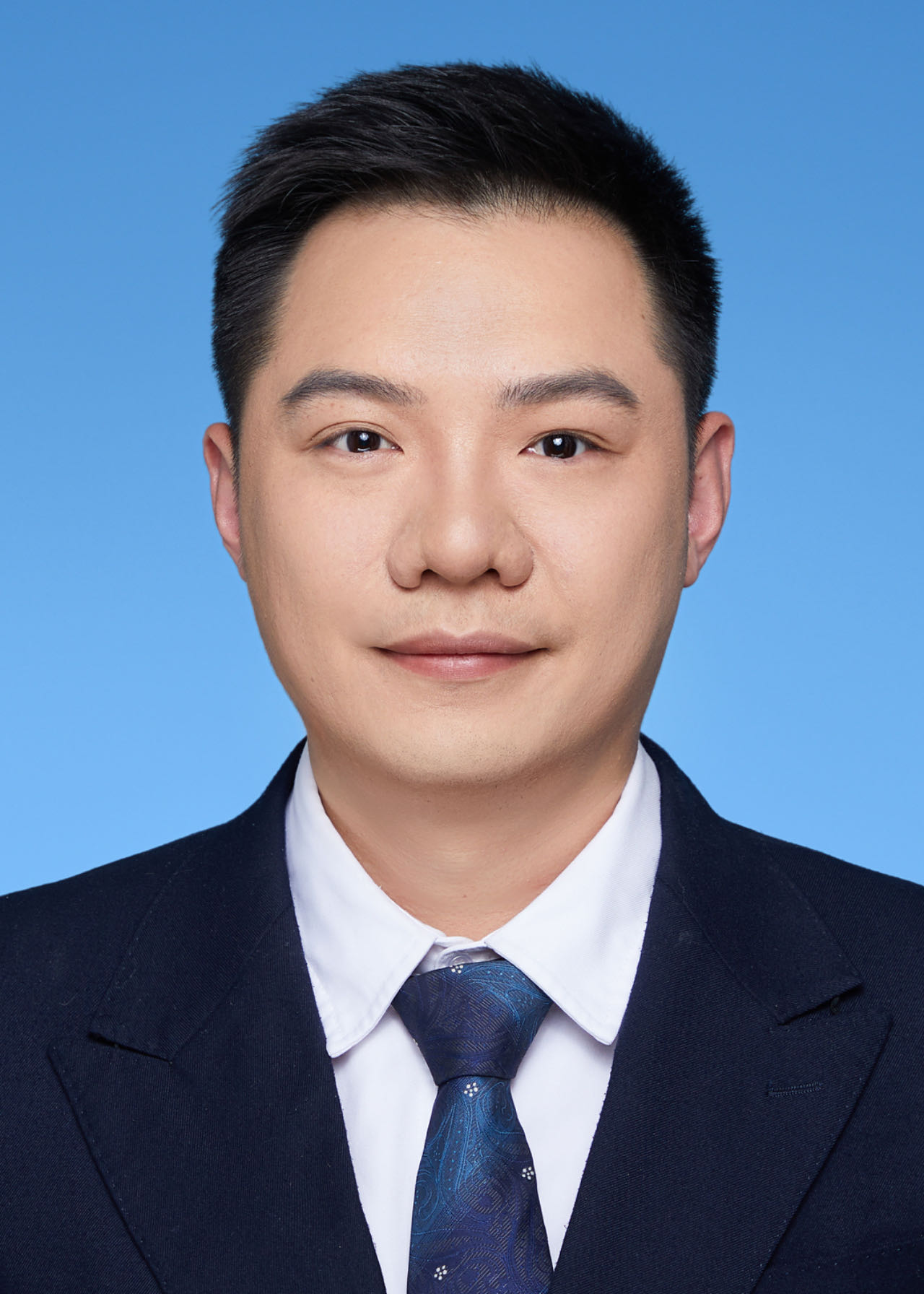}}]{Chao Zhang} is a postdoctoral researcher at Tsinghua University. He was awarded the Ph.D. degree in Computer Science at the University of Helsinki, Finland. He has given a tutorial on HTAP databases in SIGMOD 2022 and gave a tutorial on cloud databases in VLDB 2022. He serves as a PC member of SIGMOD 2024-2025, VLDB 2023 Tutorial, and ICDE 2023. His research interests focus on heterogeneous database management systems.
\end{IEEEbiography}

\vspace{-3 em}

\begin{IEEEbiography}[{\includegraphics[width=1in,height=1.25in,clip,keepaspectratio]{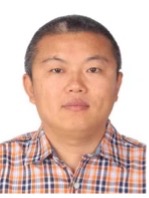}}]{Guoliang Li}
is an IEEE fellow, and a full professor at the Department of Computer Science, Tsinghua University. His research interests include database systems, large-scale data cleaning, and integration. He received the VLDB 2017 Early Research Contribution Award, TCDE 2014 Early Career Award, SIGMOD 2023 Best Papers,  VLDB 2020 Best Papers, and ICDE 2018 Best Paper. He served as a general chair of SIGMOD 2021, a demo chair of VLDB 2021, and an industry chair of ICDE 2022.
\end{IEEEbiography}

\vspace{-3 em}

\begin{IEEEbiography}[{\includegraphics[width=1in,height=1.25in,clip,keepaspectratio]{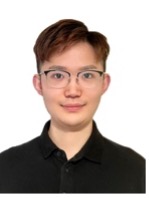}}]{Jintao Zhang} is a master student at Tsinghua University. He received his bachelor's degree in Computer Science at Xidian University. His research interests focus on the intersection between database and machine learning.
\end{IEEEbiography}

\vspace{-3 em}

\begin{IEEEbiography}[{\includegraphics[width=1in,height=1.25in,clip,keepaspectratio]{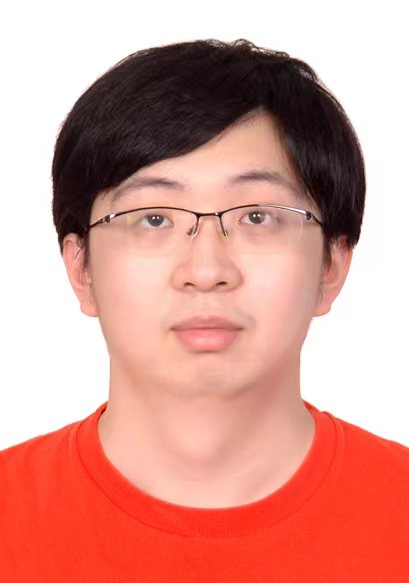}}]{Xinning Zhang} is a master student at Tsinghua University. He received his bachelor's degree in Computer Science at Zhejiang University. His research interests focus on HTAP databases.
\end{IEEEbiography}

\vspace{-3 em}

\begin{IEEEbiography}[{\includegraphics[width=1in,height=1.25in,clip,keepaspectratio]{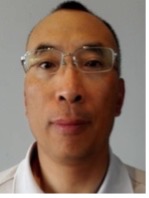}}]{Jianhua Feng}
is a full professor at the Department of Computer Science, Tsinghua University. He received his bachelor's degree in Computer Science at Tsinghua University. His research interests focus on cutting-edge database management systems.
\end{IEEEbiography}

\end{document}